\documentclass[aps,prl,reprint,twocolumn,superscriptaddress,showpacs,notitlepage]{revtex4-1}

\usepackage{amsfonts}
\usepackage{mathrsfs}
\usepackage{amsmath}
\usepackage{amssymb}
\usepackage{color}
\usepackage{graphicx}
\usepackage{bm}
\usepackage{float}
\usepackage{xspace}
\usepackage{epstopdf}
\usepackage{dcolumn}
\usepackage{longtable}
\usepackage{multirow}
\usepackage{booktabs}
\usepackage[colorlinks, linkcolor=blue,anchorcolor=blue,citecolor=blue,urlcolor=blue]{hyperref}

\usepackage{subfiles}
\usepackage{threeparttable}

\providecommand{\tabularnewline}{\\}
\newcommand{\tabincell}[2]{\begin{tabular}{@{}#1@{}}#2\end{tabular}}

\begin{document}
\title{Higher-Order Nodal Points in Two Dimensions}
\author{Weikang Wu}
\email{weikang.wu@ntu.edu.sg}
\address{Research Laboratory for Quantum Materials, Singapore University of Technology and Design, Singapore 487372, Singapore}
\address{Division of Physics and Applied Physics, School of Physical and Mathematical Sciences, Nanyang Technological University, Singapore 637371, Singapore}

\author{Ying Liu}
\thanks{W. Wu and Y. Liu contributed equally to this work.}
\address{School of Materials Science and Engineering, Hebei University of Technology, Tianjin 300130, China}
\address{Research Laboratory for Quantum Materials, Singapore University of Technology and Design, Singapore 487372, Singapore}

\author{Zhi-Ming Yu}
\address{Key Lab of Advanced Optoelectronic Quantum Architecture and Measurement (MOE), Beijing Key Lab of Nanophotonics \& Ultrafine Optoelectronic Systems, and School of Physics, Beijing Institute of Technology, Beijing 100081, China}
\address{Research Laboratory for Quantum Materials, Singapore University of Technology and Design, Singapore 487372, Singapore}

\author{Y. X. Zhao}
\affiliation{National Laboratory of Solid State Microstructures and Department of Physics, Nanjing University, Nanjing 210093, China}
\affiliation{Collaborative Innovation Center of Advanced Microstructures, Nanjing University, Nanjing 210093, China}

\author{Weibo Gao}
\address{Division of Physics and Applied Physics, School of Physical and Mathematical Sciences, Nanyang Technological University, Singapore 637371, Singapore}

\author{Shengyuan A. Yang}
\address{Research Laboratory for Quantum Materials, Singapore University of Technology and Design, Singapore 487372, Singapore}

\begin{abstract}
A two-dimensional (2D) topological semimetal is characterized by the nodal points in its low-energy band structure. While the linear nodal points have been extensively studied, especially in the context of graphene, the realm beyond linear nodal points remains largely unexplored.
Here, we explore the possibility of higher-order nodal points, i.e., points with higher-order energy dispersions, in 2D systems. We perform an exhaustive search over all 80 layer groups both with and without spin-orbit coupling (SOC), and reveal all possible higher-order nodal points. We show that they can be classified into two categories: the quadratic nodal point (QNP) and the cubic nodal point (CNP).  All the 2D higher-order nodal points have twofold degeneracy, and the order of dispersion cannot be higher than three. QNPs only exist in the absence of SOC, whereas CNPs only exist in the presence of SOC. Particularly, the CNPs represent a new topological state not known before. We show that they feature nontrivial topological charges, leading to extensive topological edge bands. Our work completely settles the problem of higher-order nodal points, discovers novel topological states in 2D, and provides detailed guidance to realize these states. Possible material candidates and experimental signatures are discussed. 
\end{abstract}
\maketitle


Topological semimetals have been attracting tremendous interest in the past decade \cite{Armitage2018Weyl-RMP,Lv2021Experimental-RMP}. These states feature protected band degeneracies, such as nodal points \cite{Murakami2007Phase-NJoP,Wan2011Topological-PRB,Xu2011Chern-PRL,Young2012Dirac-PRL,Wang2012Dirac-PRB,Wang2013Three-PRB,Yang2014Dirac-PRL,Dai2016Weyl-NP,Bradlyn2016Dirac-S,Zhu2016Triple-PRX,Chang2017Unconventional-PRL,Tang2017Multiple-PRL,Zhang2018Double-PRL,Chang2018Topological-NM}, nodal lines \cite{Burkov2011Topological-PRB,Weng2015Topological-PRB,Chen2015Nanostructured-NL,Mullen2015Line-PRL,Fang2015Topological-PRB,Yu2015Topological-PRL,Kim2015Dirac-PRL,Zhao2016Topological-PRB,Fang2016Topological-CPB}, or even nodal surfaces \cite{Zhong2016Towards-N, Liang2016Node-PRB, Wu2018Nodal-PRB}, which generate unusual low-energy quasiparticle excitations and lead to novel physical effects. A prominent example is graphene \cite{Novoselov2004Electric-S,Zhang2005Experimental-N,Novoselov2005Two-N}. Its symmetry-protected linear nodal points at Fermi level give rise to massless Dirac electrons in two-dimensions (2D), underlying many fascinating properties of graphene \cite{CastroNeto2009Electronic-RMP}.

The order of dispersion is an important character for classifying nodal points. It determines the scaling of the density of states (DOS), which enters almost all macroscopic physical properties.  In most cases, the bands cross linearly at a nodal point, so that the leading term in the band splitting $\Delta E$ is of linear order in the momentum: $\Delta E\sim k$.
Nevertheless, under certain special symmetry constraints, the linear term may be forbidden, then band splitting would start at a higher order. Such higher-order nodal points (and also nodal lines) have been systematically studied in three dimensions \cite{Fang2012Multi-PRL,Yang2014Classification-NC,Yu2019Quadratic-PRB,Wu2020Higher-PRB,Zhang2021Magnetic-PRB}. It was found that the highest dispersion is of cubic order, and all possible types were enumerated along with the protecting space groups \cite{Yu2021Encyclopedia-apa}.

In comparison, for 2D, our knowledge of higher-order nodal points is still rather primitive. The only type known so far is the so-called double Weyl point \cite{Zhu2016Blue-NL}, which has twofold degeneracy and quadratic band splitting. It was found in a few 2D materials, including the blue phosphorene oxide \cite{Zhu2016Blue-NL}, monolayer Mg$_2$C \cite{Wang2018Monolayer-PRM}, and monolayer Na$_2$O  and K$_2$O \cite{Hua2020Tunable-AS}. However, many fundamental questions are yet to be answered. For example, the double Weyl point was only reported in spinless systems, namely in the absence of spin-orbit coupling (SOC). Is it possible to stabilize it in the presence of SOC? Beyond double Weyl points, is there any other type of higher-order points, with different dispersion order or different degeneracy?

\begin{table*}[htbp]
	\centering
	\caption{QNPs in 2D systems. They only exist in the absence of SOC. HSP stands for the high-symmetry point (which is the location of the QNP). PG denotes the point group at the HSP. The IRR column gives the irreducible representation corresponding to the QNP. In the effective models, the overall energy shift term is neglected, as stated in the main text.}\label{table:wosoc}
	\scalebox{0.95}{
		\renewcommand{\arraystretch}{1.5}
		
		\begin{tabular}{ c c c c }
			\hline\hline
			LGs (HSPs) & PG & IRR & $\mathcal{H}_{\text{eff}}$ \tabularnewline
			\hline
			49 ($\Gamma$/$M$) & $C_{4}$ & $\{{^{1}E},{^{2}E}\}$ & \multirow{3}{*}{$(\alpha k_{+}^{2}+\beta k_{-}^{2})\sigma_{+}+\text{H.c.}$} \tabularnewline
			50 ($\Gamma$/$M$) & $S_{4}$ & $\{{^{1}E},{^{2}E}\}$ &  \tabularnewline
			51 ($\Gamma$/$M$), 52 ($\Gamma$) & $C_{4h}$ & $\{{^{1}E_{g}},{^{2}E_{g}}\}$ or $\{^{1}E_{u},{}^{2}E_{u}\}$ &  \tabularnewline
			\tabularnewline 
			53 ($\Gamma$/$M$), 54 ($\Gamma$) & $D_{4}$ & $E$ & \multirow{4}{*}{$(ak_{+}^{2}+bk_{-}^{2})\sigma_{+}+\text{H.c.}$} \tabularnewline
			55 ($\Gamma$/$M$), 56 ($\Gamma$) & $C_{4v}$ & $E$ &  \tabularnewline
			61 ($\Gamma$/$M$), 62 ($\Gamma$), 63 ($\Gamma$), 64 ($\Gamma$) & $D_{4h}$ & $E_{g}$ or $E_{u}$ &  \tabularnewline
			57 ($\Gamma$/$M$), 58 ($\Gamma$), 59 ($\Gamma$/$M$), 60 ($\Gamma$) & $D_{2d}$ & $E$ & \tabularnewline
			\hline
			65 ($\Gamma$) & $C_{3}$ & $\{{^{1}E},{^{2}E}\}$ & \multirow{3}{*}{$\alpha k_{+}^{2}\sigma_{+}+\text{H.c.}$} \tabularnewline
			66 ($\Gamma$) & $C_{3i}$ & $\{{^{1}E_{g}},{^{2}E_{g}}\}$ or $\{^{1}E_{u},{}^{2}E_{u}\}$ &  \tabularnewline
			74 ($\Gamma$) & $C_{3h}$ & $\{{^{1}E'},{^{2}E'}\}$ or $\{{^{1}E''},{^{2}E''}\}$ & \tabularnewline
			\tabularnewline 
			67 ($\Gamma$), 68 ($\Gamma$) & $D_{3}$ & $E$ & \multirow{4}{*}{$ak_{+}^{2}\sigma_{+}+\text{H.c.}$} \tabularnewline
			69 ($\Gamma$), 70 ($\Gamma$) & $C_{3v}$ & $E$ &  \tabularnewline
			71 ($\Gamma$), 72 ($\Gamma$) & $D_{3d}$ & $E_{g}$ or $E_{u}$ &  \tabularnewline
			78 ($\Gamma$), 79 ($\Gamma$) & $D_{3h}$ & $E'$ or $E''$ &  \tabularnewline
			\hline
			73 ($\Gamma$) & $C_{6}$ & $\{{^{1}E_{1}},{^{2}E_{1}}\}$ or $\{{^{1}E_{2}},{^{2}E_{2}}\}$ & \multirow{2}{*}{$\alpha k_{+}^{2}\sigma_{+}+\text{H.c.}$} \tabularnewline
			75 ($\Gamma$) & $C_{6h}$ & \tabincell{c}{$\{{^{1}E_{1g}},{^{2}E_{1g}}\}$ or $\{{^{1}E_{1u}},{^{2}E_{1u}}\}$ or $\{{^{1}E_{2g}},{^{2}E_{2g}}\}$ or $\{{^{1}E_{2u}},{^{2}E_{2u}}\}$} &  \tabularnewline
			\tabularnewline 
			76 ($\Gamma$) & $D_{6}$ & $E_{1}$ or $E_{2}$ & \multirow{3}{*}{$ak_{+}^{2}\sigma_{+}+\text{H.c.}$} \tabularnewline
			77 ($\Gamma$) & $C_{6v}$ & $E_{1}$ or $E_{2}$ &  \tabularnewline
			80 ($\Gamma$) & $D_{6h}$ & $E_{1g}$ or $E_{1u}$ or $E_{2g}$ or $E_{2u}$ &  \tabularnewline
			\hline\hline
		\end{tabular}
	}
\end{table*}

In this work, we fully address the above questions by a thorough study of higher-order nodal points in 2D. Via a systematic analysis of the 80 layer groups (LGs) with the time reversal symmetry, both in the absence and presence of SOC, we obtain all possible stable 2D higher-order nodal points. The results are listed in Table~\ref{table:wosoc} and \ref{table:wsoc}. Our key findings are the following. (i) We find that higher-order nodal points in 2D can only have twofold degeneracy. In other words, they are ``Weyl" in terms of degeneracy. (ii) Beyond the quadratic order, we reveal a previously unknown possibility --- the cubic nodal point (CNP), and moreover, it is the only other possibility. (iii) The quadratic nodal point (QNP) only exists in the absence of SOC, whereas the CNP only exists in the presence of SOC. The layer groups, the locations in the Brillouin zone (BZ), the symmetry representations, and the effective models for these nodal points are obtained and presented in Table~\ref{table:wosoc} and \ref{table:wsoc}.
In addition, we also discuss their topological charge, the edge states, and measurable physical signatures, such as magnetoresponse, the (joint) density of states, and distincted Landau levels. Our work discovers a new topological state --- the 2D CNP semimetal, completely settles the exploration of 2D higher-order nodal points, and provides detailed guidance for searching/engineering such topological states in real materials and artificial systems.

\begin{table*}[htbp]
	\centering
	\caption{CNPs in 2D systems. They only exist in the presence of SOC. The arrangement is similar to Table \ref{table:wosoc}. }\label{table:wsoc}
	\scalebox{1.0}{
			\renewcommand{\arraystretch}{1.5}
			\begin{tabular}{ c c c c c c }
				\hline\hline
				LGs (HSPs) & PG & IRR & $\mathcal{H}_{\text{eff}}$ \tabularnewline
				\hline
				65 ($\Gamma$) & $C_{3}$ & $\{\bar{E},\bar{E}\}$ & $[(\alpha k_{+}^{3}+\beta k_{-}^{3})\sigma_{+} +\text{H.c.}] + (\gamma k_{+}^{3} +\gamma^{*} k_{-}^{3})\sigma_{z}$ \tabularnewline
				67 ($\Gamma$), 68 ($\Gamma$) & $D_{3}$ & $\{{^{1}\bar{E}},{^{2}\bar{E}}\}$ & \multirow{2}{*}{$[\alpha(k_{+}^{3}+k_{-}^{3})\sigma_{+} +\text{H.c.}] + ic(k_{+}^{3} - k_{-}^{3})\sigma_{z}$\tnote{1}} \tabularnewline
				69 ($\Gamma$), 70 ($\Gamma$) & $C_{3v}$ & $\{{^{1}\bar{E}},{^{2}\bar{E}}\}$ &  \tabularnewline
				\hline
				73 ($\Gamma$) & $C_{6}$ & $\{{^{1}\bar{E}_{1}},{^{2}\bar{E}_{1}}\}$ & $(\alpha k_{+}^{3}+\beta k_{-}^{3})\sigma_{+}+\text{H.c.}$ \tabularnewline
				76 ($\Gamma$) & $D_{6}$ & $\bar{E}_{3}$ & \multirow{2}{*}{$(ak_{+}^{3}+bk_{-}^{3})\sigma_{+}+\text{H.c.}$} \tabularnewline
				77 ($\Gamma$) & $C_{6v}$ & $\bar{E}_{3}$ &  \tabularnewline
				\hline\hline
			\end{tabular}
			
			
	}
\end{table*}

\textit{\textcolor{blue}{Approach.}}\label{sec:approach}
We first describe our approach to obtain all possible higher-order nodal points. As mentioned, the higher-order nodal points must require certain symmetries to (1) protect its stability and (2) eliminate the low-order terms in the band splitting. Thus, such points must reside at high-symmetry points or high-symmetry paths in the 2D Brillouin zone (BZ). Therefore, our procedure is to scan the high-symmetry points and paths for each LG. A nodal point would correspond to a higher-dimensional irreducible representation (IRR) ($d\geq 2$) of the little group at a high-symmetry point, or a crossing between two bands with different IRRs on a high-symmetry path. For systems with/without SOC, single/double valued representations should be adopted. These IRRs for the layer groups can be obtained from the standard reference \cite{Bradley2009Mathematical-OUP}. Then, for each nodal point, we construct its $k\cdot p$ effective model based on the IRRs, from which the order of dispersion can be read off.
This procedure is applied to all 80 LGs with and without SOC \cite{SupplementaryMaterial}, and the results are presented in Table~\ref{table:wosoc} and \ref{table:wsoc}.

Here are some general observations. First, beyond QNPs, we discover a new possibility --- the CNPs, which represents a previously unknown topological structure. And our result shows that these two are the only possible types of higher-order nodal points in 2D.

Second, all possible higher-order points have twofold degeneracy. The generic $k\cdot p$ effective model expanded at such a point can be expressed as
\begin{equation}\label{Heff}
  \mathcal{H}_\text{eff}=\varepsilon(\bm k)+f(\bm k)\sigma_+ +f^*(\bm k)\sigma_- +g(\bm k)\sigma_z,
\end{equation}
where the momentum $\bm k$ and the energy are measured from the nodal point, $\sigma_{\pm}=\frac{1}{2}(\sigma_{x}\pm i\sigma_{y})$, and the $\sigma_i$'s are the Pauli matrices. $\varepsilon(\bm{k})$ is an overall energy shift, which does not affect the band splitting, hence it will be omitted hereafter.  $f $ and $g$ are functions of $\bm k$, with $f=g=0$ at the nodal point $\bm k=0$. $f$ is generally complex, whereas $g$ must be real. The order of the point is determined by leading order terms in $f$ and $g$.

Third, the QNPs only exist in the absence of SOC, which hence just correspond to the double Weyl points in Refs. \cite{Zhu2016Blue-NL,Wang2018Monolayer-PRM,Hua2020Tunable-AS}. There is no QNP in 2D systems with SOC. In sharp contrast, the CNPs show the opposite behavior, i.e., they only exist in the presence of SOC.

Finally, all possible higher-order points are found at high-symmetry points of BZ. For QNPs, they mostly occur at $\Gamma$, and in a few cases with fourfold axis, they can also appear at the $M$ $(\pi,\pi)$ point. In comparison, all CNPs occur at the $\Gamma$ point.
\

\textit{\textcolor{blue}{Quadratic nodal point.}}\label{sec:spinless_QWP}
In {Table \ref{table:wosoc}}, we present all possible QNPs with their layer groups, point groups, IRRs, and effective models.

To see how the quadratic dispersion arises from symmetry, let us consider LG~65 ($p3$) as an example. the little co-group at $\Gamma$ is generated by the threefold rotation $C_{3z}$ and time reversal symmetry $\mathcal{T}$, which can be represented as
\begin{equation}
C_{3z} = e^{i\frac{2\pi}{3}\sigma_{z}}, \qquad \mathcal{T} = \sigma_{x}\mathcal{K}.
\end{equation}
in the degenerate basis $\{{^{1}E},{^{2}E}\}$ corresponding to the QNP. The symmetry constraints on the effective model are given by
\begin{subequations}\label{eq:SymmConstrQWP2}
	\begin{eqnarray}
	C_{3z}\mathcal{H}_{\text{eff}}^{65}(\bm{k})C_{3z}^{-1} & = & \mathcal{H}_{\text{eff}}^{65}(R_{3z}\bm{k}), \label{eq:SymmConstrQWP2_C3} \\
	\mathcal{T}\mathcal{H}_{\text{eff}}^{65}(\bm{k})\mathcal{T}^{-1} & = & \mathcal{H}_{\text{eff}}^{65}(-\bm k), \label{eq:SymmConstrQWP2_T}
	\end{eqnarray}
\end{subequations}
where $R_{3z}$ is the threefold rotation acting on $\bm{k}$. Substituting the general form Eq.~(\ref{Heff}) into these equations, one finds the following constraints on $f$ and $g$: {$f(\bm{k})=f(-\bm{k})=e^{i\frac{2\pi}{3}}f(R_{3z}\bm{k})$ and $g(\bm{k})=-g(-\bm{k})=g(R_{3z}\bm{k})$.}
Thus, linear terms in $f$ and $g$ are forbidden. Up to the $k^2$ order, the effective model for the QNP in LG~65 takes the form of
\begin{equation}\label{eq:HamQWP_p3}
\mathcal{H}_{\text{eff}}^{65}(\bm{k}) = \alpha k_{+}^{2}\sigma_{+} + \text{H.c.},
\end{equation}
where the Greek letter $\alpha$ stands for a complex parameter. Note that the $k^2$ term in $g$ is also eliminated by $\mathcal{T}$, hence $g$ starts at the cubic order. It follows that this QNP also enjoys an approximate chiral symmetry $\{\mathcal{S}, \mathcal{H}_{\text{eff}}^{65}\}=0$, with $\mathcal{S}=\sigma_z$.

Similarly, one can expect to find QNPs in the supergroups of LG~65. The extra symmetries will give additional constraints on the QNP model. For example, adding a vertical mirror to LG~65 expands it to LG~69 or 70 (depending on the mirror orientation), and the mirror further requires the parameter $\alpha$ in (\ref{eq:HamQWP_p3}) to be real. In {Table \ref{table:wosoc}}, we denote real parameters by roman letters.

On the other hand, adding a $C_{2z}$ symmetry expands LG~65 to LG~73, with a sixfold axis. This extra symmetry completely suppresses the $g$ term to all orders. Thus, the emergent chiral symmetry $\mathcal{S}$ becomes more robust for the QNP on a sixfold axis.

We have also verified the results in {Table \ref{table:wosoc}} by constructing explicit lattice models for the candidate LGs. An example for LG~65 is shown in the Supplemental Material (SM) \cite{SupplementaryMaterial}, which confirms a 2D semimetal state with a QNP being the single Fermi point.

\textit{\textcolor{blue}{Cubic nodal point.}}\label{sec:spinful_CWP}
CNPs only occur in systems with SOC. From {Table \ref{table:wsoc}}, one observes that they are compatible with $C_{3z}$ or $C_{6z}$ symmetry. The effective models are derived similar to the above example, so we will not repeat here. Instead, we construct a lattice model to directly visualize the CNP.

\begin{figure}[tbp]
	\centering
	\includegraphics[width=0.48\textwidth]{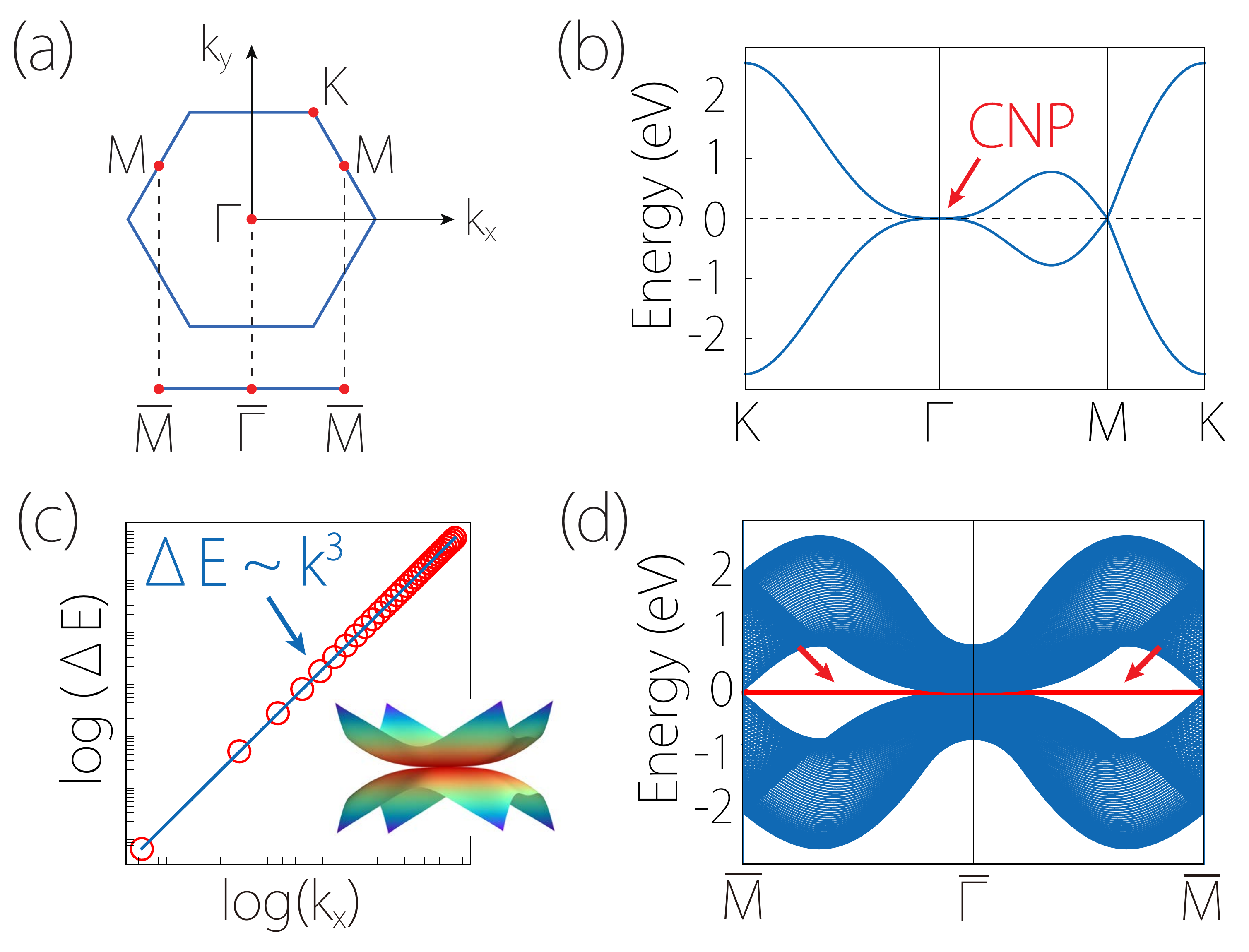}
	\caption{Lattice model with CNP. 
		(a) Brillouin zone for LG~77, along with the edge BZ for the edge normal to the $y$ direction. (b) Band structure for the lattice model in Eq.~(\ref{TB}). A CNP is found at $\Gamma$. The crossing at $M$ is a linear nodal point. (c) The log-log plot for the energy splitting around the CNP, demonstrating the leading order cubic dispersion. The inset shows the 2D band structure around the CNP. (d) Spectrum for a ribbon geometry with edges normal to $y$. The arrows indicate the topological edge band. In the figure, the model parameters are set to $t_1 = t_2 = 0$, $t_1^{\text{SO}} = 1.0$, and $t_2^{\text{SO}} = 0.3$.}
	\label{fig:LG77_wsoc}
\end{figure}

Let us consider a model for LG~77. We take a {triangular} lattice, with the active site occupying the $1a$ Wyckoff position within a unit cell. Each active site has an $s$-like orbital with two spin states, corresponding to the $\bar{E}_{3}$ IRR of the $C_{6v}$ group. Under this basis, the generators of LG~77 can be represented as
\begin{equation}
C_{6z} = i \sigma_{z},\quad
M_{x} = i \sigma_{x},\quad
\mathcal{T} = -i \sigma_{y}\mathcal{K},
\end{equation}
where $\sigma_i$ here act on the spin space. Then, following the method in {Ref.~\cite{Wieder2016Spin-PRB,Yu2019Circumventing-PRB}}, a lattice model respecting these symmetries can be obtained as
\begin{widetext}
  \begin{equation}
\begin{aligned}\label{TB}
H_{\text{latt}}^{\text{LG77}} = & t_1\Big(\cos k_x + 2\cos\frac{k_x}{2}\cos\frac{\sqrt{3}k_y}{2}\Big) + t_2\Big(\cos\sqrt{3}k_y + 2\cos\frac{3k_x}{2} \cos\frac{\sqrt{3}k_y}{2}\Big) \\
&+ t_1^{\text{SO}}\Big(\sin k_x - 2 \sin\frac{k_x}{2} \cos\frac{\sqrt{3}k_y}{2}\Big)\sigma_y
+ t_2^{\text{SO}}\Big(-\sin\sqrt{3}k_y + \sin\frac{3k_x}{2}\cos\frac{\sqrt{3}k_y}{2}\Big)\sigma_x,
\end{aligned}
\end{equation}
\end{widetext}
where $t$'s are hopping amplitudes, and the last two terms are symmetry-allowed SOC terms. Evidently, the first two terms in (\ref{TB}) represent an overall energy shift, hence they are inessential for the emergence of CNP. One can readily verify that expanding this model at $\Gamma$ recovers the CNP effective model shown in Table \ref{table:wsoc}.
Figure \ref{fig:LG77_wsoc}(b) shows a typical band structure obtained for this model. The CNP at $\Gamma$ can be clearly observed and its cubic dispersion is explicitly confirmed in Fig. \ref{fig:LG77_wsoc}(c).

It should be noted that in Fig.~\ref{fig:LG77_wsoc}(b), besides the CNP, there also exist three linear nodal points at the three $M$ points. This is not accidental. Below, we will see that it is a consequence of the nontrivial topological charge of the CNP.

\


\textit{\textcolor{blue}{Topological charge.}}\label{sec:topological_charge}
A nodal point in 2D can be characterized by the Berry phase $\gamma_{C}$ defined on a small circle $C$ enclosing the point,
\begin{equation}
\gamma_{C} = \oint_{C} \bm{\mathcal{A}}(\bm{k})\cdot d\bm{k} \quad \mod 2\pi,
\end{equation}
where $\bm{\mathcal{A}}$ is the Berry connection for the occupied states. The linear Weyl point in graphene has $\gamma_C=\pi$. One can easily see that
QNP and CNP will have $\gamma_C=0$ and $\pi$, respectively. It should be noted that under certain symmetries, e.g., $\mathcal{PT}$ or $C_{2z}\mathcal{T}$ in the absence of SOC, and $C_{2z}\mathcal{T}$ in the presence of SOC, the $\pi$-quantization of $\gamma_C$ becomes exact. This applies to most LGs in Table~\ref{table:wosoc} and the latter three LGs in Table~\ref{table:wsoc}. $\gamma_C$ thus defines
 a $\mathbb{Z}_2$-valued topological charge for the nodal points.

Furthermore, all QNPs in Table~\ref{table:wosoc} and the CNPs of the latter three LGs in Table~\ref{table:wsoc} have an emergent chiral symmetry $\mathcal{S}=\sigma_z$, as can be seen from their effective models $\mathcal{H}_\text{eff}$. Under this symmetry, we actually can define a $\mathbb{Z}$-valued topological charge on the small circle $C$, given by the 1D winding number \cite{Schnyder2008Classification-PRB}
\begin{equation}
\mathcal{N}_{C} = \frac{1}{4\pi i}\oint_{C} \text{Tr}\ \sigma_{z} \mathcal{H}_{\text{eff}}^{-1}(\bm{k}) \nabla_{\bm{k}} \mathcal{H}_{\text{eff}}(\bm{k}) \cdot d\bm{k}.
\end{equation}
One can check that QNP and CNP have $|\mathcal{N}_{C}| = 2$ and $ 3$, respectively. And the relation $(\gamma_C/\pi)=\mathcal{N}_{C} \mod 2$ holds between the two charges.

This explains the necessary appearance of extra linear Weyl points in Fig. \ref{fig:LG77_wsoc}(b). Because the CNP has a nontrivial Berry phase $\gamma_C=\pi$, the no-go theorem dictates that it cannot exist as the single Fermi point but must be accompanied by an odd number of linear Weyl points \cite{Zhao2016Novel-PRL}.

The topological charge is also connected to the edge states. The nontrivial $\gamma_C=\pi$ for a CNP means that there will be topological edge band at a generic edge. For example, in Fig.~\ref{fig:LG77_wsoc}(d), we consider the edge of our lattice model (\ref{TB}) normal to the $y$ direction. One observes that the edge band spans over the whole edge BZ, connecting the edge projection of the CNP and the linear Weyl points.

\

\textit{\textcolor{blue}{Discussion.}}\label{sec:experimental_signature}
We have exhaustively explored all possible higher-order nodal points in 2D and uncovered novel 2D topological semimetal states with CNPs. The symmetry conditions provided here will facilitate the search/design of real materials as well as artificial systems to achieve the proposed novel states.

Combining the symmetry conditions with \emph{ab-initio} calculations, we have predicted a few 2D material candidates, including QNPs in monolayer Ti$_3$C$_2$F$_2$ and V$_2$TiC$_2$F$_2$, and CNP in monolayer Bi$_2$O (see Supplemental Material) \cite{SupplementaryMaterial}.

Experimentally, these nodal points and the topological edge states can be directly probed by angle-resolved photoemission spectroscopy (ARPES) \cite{Ohta2006Controlling-S,Feng2017Experimental-NC,Collins2018Electric-N} and scanning tunneling spectroscopy (STS) \cite{Zheng2016Atomic-AN,Zhu2018Quasiparticle-NC}. The distinct Berry phase $\gamma_C$ for QNP and CNP can manifest in a range of effects \cite{Mikitik1999Manifestation-PRL,Shoenberg2009Magnetic-CUP}. For example, it would result in weak (anti-)localization for QNP (CNP), and it enters the phase offset in the Shubnikov-de Haas oscillation $\delta\rho_{xx}\sim
\cos[2\pi(\frac{F}{B} -  \frac{1}{2} + \frac{\gamma_{C}}{2\pi})]$. The distinct leading order dispersion leads to distinct scaling in the DOS. Particularly, the joint DOS for CNP (QNP)  scales as  $\sim E^{-1/3}$ ($\sim E^0$), which impacts the optical interband transitions. The order of dispersion also affects the Landau level scaling: the energy for the $n$-th Landau level would scale as $\sim (nB)^{3/2}$ ($\sim nB$) for a CNP (QNP). Here, it might appear that the case with QNP is nothing different from a conventional 2D electron gas. However, there exists a distinction due to nontrivial the topological charge $\mathcal{N}_C$: For 2D electron gas, there is no zero-energy Landau level, the first level occurs at $\hbar eB/(2m^*)$, whereas for QNP and CNP, there must be $|\mathcal{N}_C|$ zero Landau levels, as dictated by the index theorem \cite{Zhao2021Index-PRL}.

\end{document}


\title{Supplementary Materials for ``Higher-Order Nodal Points in Two Dimensions''}

\author{Weikang Wu}
\email{weikang.wu@ntu.edu.sg}
\address{Research Laboratory for Quantum Materials, Singapore University of Technology and Design, Singapore 487372, Singapore}
\address{Division of Physics and Applied Physics, School of Physical and Mathematical Sciences, Nanyang Technological University, Singapore 637371, Singapore}

\author{Ying Liu}
\thanks{W. Wu and Y. Liu contributed equally to this work.}
\address{School of Materials Science and Engineering, Hebei University of Technology, Tianjin 300130, China}
\address{Research Laboratory for Quantum Materials, Singapore University of Technology and Design, Singapore 487372, Singapore}

\author{Zhi-Ming Yu}
\address{Key Lab of Advanced Optoelectronic Quantum Architecture and Measurement (MOE), Beijing Key Lab of Nanophotonics \& Ultrafine Optoelectronic Systems, and School of Physics, Beijing Institute of Technology, Beijing 100081, China}
\address{Research Laboratory for Quantum Materials, Singapore University of Technology and Design, Singapore 487372, Singapore}

\author{Y. X. Zhao}
\affiliation{National Laboratory of Solid State Microstructures and Department of Physics, Nanjing University, Nanjing 210093, China}
\affiliation{Collaborative Innovation Center of Advanced Microstructures, Nanjing University, Nanjing 210093, China}

\author{Weibo Gao}
\address{Division of Physics and Applied Physics, School of Physical and Mathematical Sciences, Nanyang Technological University, Singapore 637371, Singapore}

\author{Shengyuan A. Yang}
\address{Research Laboratory for Quantum Materials, Singapore University of Technology and Design, Singapore 487372, Singapore}

\date{\today}
\maketitle

\section{Layer group and space group}
For a three-dimensional (3D) crystal, its space group (SG) $G$ involves the lattice translations along three directions. The corresponding lattice translation group can be denoted as $T_{1}$, $T_{2}$, $T_{3}$, respectively. If one of translation groups is violated, the SG would be reduced to a layer group (LG) which describes the symmetry of a two-dimensional (2D) material. This means that each layer group should be isomorphic to a factor group $G/T_{1}$ and one can build a correspondence between SGs and LGs as shown in Table~\ref{SM:table_LG} \cite{Po2017Symmetry-NC}. Then, the irreducible representations (IRRs) for a LG can be deduced from the corresponding SG whose IRRs can be obtained from the standard reference \cite{Bradley2009Mathematical-OUP}. This serves as the foundation of searching for all possible higher-order nodal points in 2D.

\begin{table}[b]
	\centering
	\caption{Correspondence between layer groups and space groups. Here, LG and SG denote layer group and space group, respectively.}\label{SM:table_LG}
	\centering
	\scalebox{0.88}{
	\renewcommand{\arraystretch}{1.5}
	\begin{tabular}{ c c c c|c c c c|c c c c|c c c c }
		\hline\hline 
		LG No. & Symbol & SG No. & Symbol & LG No. & Symbol & SG No. & Symbol & LG No. & Symbol & SG No. & Symbol & LG No. & Symbol & SG No. & Symbol\tabularnewline
		\hline 
		1 & $p1$ & 1 & $P1$ & 21 & $p2_{1}2_{1}2$ & 18 & $P2_{1}2_{1}2$ & 41 & $pmma$ & 51 & $Pmma$ & 61 & $p4/mmm$ & 123 & $P4/mmm$\tabularnewline
		2 & $p\bar{1}$ & 2 & $P\bar{1}$ & 22 & $c222$ & 21 & $C222$ & 42 & $pman$ & 53 & $Pmna$ & 62 & $p4/nbm$ & 125 & $P4/nbm$\tabularnewline
		3 & $p112$ & 3 & $P2$ & 23 & $pmm2$ & 25 & $Pmm2$ & 43 & $pbaa$ & 54 & $Pcca$ & 63 & $p4/mbm$ & 127 & $P4/mbm$\tabularnewline
		4 & $p11m$ & 6 & $Pm$ & 24 & $pma2$ & 28 & $Pma2$ & 44 & $pbam$ & 55 & $Pbam$ & 64 & $p4/nmm$ & $129$ & $P4/nmm$\tabularnewline
		5 & $p11a$ & 7 & $Pc$ & 25 & $pba2$ & 32 & $Pba2$ & 45 & $pbma$ & 57 & $Pbcm$ & 65 & $p3$ & 143 & $P3$\tabularnewline
		6 & $p112/m$ & 10 & $P2/m$ & 26 & $cmm2$ & 35 & $Cmm2$ & 46 & $pmmn$ & 59 & $Pmmn$ & 66 & $p\bar{3}$ & 147 & $P\bar{3}$\tabularnewline
		7 & $p112/a$ & 13 & $P2/c$ & 27 & $pm2m$ & 25 & $Pmm2$ & 47 & $cmmm$ & 65 & $Cmmm$ & 67 & $p312$ & 149 & $P312$\tabularnewline
		8 & $p211$ & 3 & $P2$ & 28 & $pm2_{1}b$ & 26 & $Pmc2_{1}$ & 48 & $cmme$ & 67 & $Cmme$ & 68 & $p321$ & 150 & $P321$\tabularnewline
		9 & $p2_{1}11$ & 4 & $P2_{1}$ & 29 & $pb2_{1}m$ & 26 & $Pmc2_{1}$ & 49 & $p4$ & 75 & $P4$ & 69 & $p3m1$ & 156 & $P3m1$\tabularnewline
		10 & $c211$ & 5 & $C2$ & 30 & $pb2b$ & 27 & $Pcc2$ & 50 & $p\bar{4}$ & 81 & $P\bar{4}$ & 70 & $p31m$ & 157 & $P31m$\tabularnewline
		11 & $pm11$ & 6 & $Pm$ & 31 & $pm2a$ & 28 & $Pma2$ & 51 & $p4/m$ & 83 & $P4/m$ & 71 & $p\bar{3}1m$ & 162 & $P\bar{3}1m$\tabularnewline
		12 & $pb11$ & 7 & $Pc$ & 32 & $pm2_{1}n$ & 31 & $Pmn2_{1}$ & 52 & $p4/n$ & 85 & $P4/n$ & 72 & $p\bar{3}m1$ & 164 & $P\bar{3}m1$\tabularnewline
		13 & $cm11$ & 8 & $Cm$ & 33 & $pb2_{1}a$ & 29 & $Pca2_{1}$ & 53 & $p422$ & 89 & $P422$ & 73 & $p6$ & 168 & $P6$\tabularnewline
		14 & $p2/m11$ & 10 & $P2/m$ & 34 & $pb2n$ & 30 & $Pnc2$ & 54 & $p42_{1}2$ & 90 & $P42_{1}2$ & 74 & $p\bar{6}$ & 174 & $P\bar{6}$\tabularnewline
		15 & $p2_{1}/m11$ & 11 & $P2_{1}/m$ & 35 & $cm2m$ & 38 & $Amm2$ & 55 & $p4mm$ & 99 & $P4mm$ & 75 & $p6/m$ & 175 & $P6/m$\tabularnewline
		16 & $p2/b11$ & 13 & $P2/c$ & 36 & $cm2e$ & 39 & $Aem2$ & 56 & $p4bm$ & 100 & $P4bm$ & 76 & $p622$ & 177 & $P622$\tabularnewline
		17 & $p2_{1}/b11$ & 14 & $P2_{1}/c$ & 37 & $pmmm$ & 47 & $Pmmm$ & 57 & $p\bar{4}2m$ & 111 & $P\bar{4}2m$ & 77 & $p6mm$ & 183 & $P6mm$\tabularnewline
		18 & $c2/m11$ & 12 & $C2/m$ & 38 & $pmaa$ & 49 & $Pccm$ & 58 & $p\bar{4}2_{1}m$ & 113 & $P\bar{4}2_{1}m$ & 78 & $p\bar{6}m2$ & 187 & $P\bar{6}m2$\tabularnewline
		19 & $p222$ & 16 & $P222$ & 39 & $pban$ & 50 & $Pban$ & 59 & $p\bar{4}m2$ & 115 & $P\bar{4}m2$ & 79 & $p\bar{6}2m$ & 189 & $P\bar{6}2m$\tabularnewline
		20 & $p2_{1}22$ & 17 & $P222_{1}$ & 40 & $pmam$ & 51 & $Pmma$ & 60 & $p\bar{4}b2$ & 117 & $P\bar{4}b2$ & 80 & $p6/mmm$ & 191 & $P6/mmm$\tabularnewline
		\hline\hline
	\end{tabular}
}
\end{table}

\section{Derivation of effective Hamiltonian}\label{SM:sec_kdotp}
In this section, we present the detailed derivation of the effective Hamiltonians for the higher-order nodal points discussed in the main text. Since all possible higher-order points have twofold degenercy, the general form of effective Hamiltonian around such a point can be expressed as a $2 \times 2$ matrix,
\begin{equation}\label{SM:eq_Heff}
    \mathcal{H}_{\text{eff}}(\bm{k}) = \epsilon(\bm{k}) + f(\bm{k})\sigma_{+} + f^{*}(\bm{k})\sigma_{-} + g(\bm{k})\sigma_{z}
\end{equation}
where $\bm{k}=(k_{x},k_{y})$ is measured from the nodal point and $\sigma_{\pm}=\frac{1}{2}(\sigma_{x}\pm i\sigma_{y})$. Here, $\epsilon(\bm{k})$ is an overall energy shift, $f$ is a complex function and $g$ is a real function. The order of the point is determined by the leading order terms in $f$ and $g$

\subsection{Quadratic nodal points in the absence of spin-orbit coupling}\label{SM:sec_kdotp_QNPwosoc}
In the absence of spin-orbit coupling (SOC), the quadratic nodal points (QNPs) can be dictated by threefold, fourfold, and sixfold rotation or roto-inversion symmetries. 

\

\textit{\textcolor{blue}{QNPs dictated by fourfold rotation or roto-inversion symmetries ---}}\label{SM:sec_kdotp_QNPwosoc_C4}
A QNP can be guaranteed by fourfold rotation. Take the LG 49 ($p4$) as an example. The QNP is located at $\Gamma$ $(0,0)$ [or $M$ $(\pi,\pi)$], whose little co-group is $C_{4}$. The generators of $C_{4}$ include $C_{4z} \equiv C_{4z}^{+}$. In the basis of the IRR $\{^{1}E,~^{2}E\}$ \cite{Koster1963Properties-MP,Bradley2009Mathematical-OUP}, e.g. $\{x+iy, x-iy\}$, the generator and time-reversal $\mathcal{T}$ are represented by
\begin{equation}
C_{4z} = i\sigma_{z}, \qquad \mathcal{T} = \sigma_{x}\mathcal{K}
\end{equation}
where $\sigma_{i}~(i=x,y,z)$ are the Pauli matrices, and $\mathcal{K}$ denotes the complex conjugate. Under the symmetry constraints
\begin{subequations}\begin{eqnarray}
C_{4z}\mathcal{H}_{\text{eff}}(\bm{k})C_{4z}^{-1} &=& \mathcal{H}_{\text{eff}}(-k_{y},k_{x}), \\
\mathcal{T}\mathcal{H}_{\text{eff}}(\bm{k})\mathcal{T}^{-1} &=& \mathcal{H}_{\text{eff}}(-k_{x},-k_{y}),
\end{eqnarray}\end{subequations}
the effective Hamiltonian (up to second order) reads
\begin{equation}
\mathcal{H}_{\text{eff}}^{49,\ \Gamma/M}(\bm{k}) = \epsilon(\bm{k}) + [(\alpha k_{+}^{2} + \beta k_{-}^{2})\sigma_{+} + \text{H.c.}]
\end{equation}
where $\epsilon(\bm{k}) = \omega_{0}+\omega_{1}(k_{x}^{2}+k_{y}^{2})$ with $\omega_{i}$ being real parameters, $k_{\pm} = k_{x}\pm ik_{y}$, $\sigma_{\pm}=(\sigma_{x}\pm i\sigma_{y})/2$, and $\alpha$, $\beta$ are complex parameters. This effective model confirms that the nodal point is a QNP with quadratic splitting in the $k_{x}$-$k_{y}$ plane.

By adding new symmetry operations, one can construct the supergroup of LG 49, which should only change the form of Hamiltonian but if the QNP still exists, its dispersion should be still quadratic unless the second-order term is eliminated. In Table~\ref{SM:table_C4}, we list all possible supergroups of LG 49 and present the corresponding effective Hamiltonian of QNPs.

The QNPs can also be dominated by fourfold rotoinversion $S_{4z}\equiv\mathcal{P}C_{4z}^{-}$. Taking the LG 50 ($p\bar{4}$) as an example, one notes that $S_{4z}$ shares the same matrix representation to $C_{4z}$ under the same basis (e.g. $\{x+iy, x-iy\}$) at $\Gamma$ $(0,0)$ [or $M$ $(\pi,\pi)$], and they both map $(k_{x}, k_{y})$ to $(-k_{y}, k_{x})$ in the momentum space. Therefore, the effective Hamiltonian for LG 50 would have the same form to that of LG 49, which describes a QNP. One can further find possible QNPs in the supergroup of LG 50. In Table~\ref{SM:table_C4}, we list all possible supergroups of LG 50 and present the corresponding effective Hamiltonian of QNPs.

\begin{table}[t]
	\centering
	\caption{Effective Hamilonian up to third order for supergroups of the layer groups $p4$ (No. 49) and $p\bar{4}$ (No. 50) in the absence of spin-orbit coupling. Here, $S_{4z}\equiv \mathcal{P}C_{4z}^{-}$. LG, HSP, PG, IRR stand for the layer group, the high-symmetry points (HSPs), the little co-groups of HSPs, and the irreducible representation, respectively. $f(\bm{k})$ and $g(\bm{k})$ are functions given in Eq.~\ref{SM:eq_Heff}.}\label{SM:table_C4}
	\scalebox{1.0}{
	\renewcommand{\arraystretch}{1.5}
	\begin{tabular}{ c c|c|c|c|c|c|c|c }
		\hline \hline 
		No. & LG & HSP & Generators & PG & IRR & Matrix Representation & $f(\bm{k})$ & $g(\bm{k})$\tabularnewline
		\hline 
		49 & $p4$ & $\Gamma$, $M$ & $\{C_{4z}^{+}|00\}$, $\mathcal{T}$ & $C_{4}$ & $\{^{1}E,^{2}E\}$ & $C_{4z}=i\sigma_{z}$, $\mathcal{T}=\sigma_{x}\mathcal{K}$ & \multirow{6}{*}{$\alpha k_{+}^{2}+\beta k_{-}^{2}$} & \multirow{6}{*}{0}\tabularnewline
		\cline{1-7}
		50 & $p\bar{4}$ & $\Gamma$, $M$ & $\{S_{4z}^{+}|00\}$, $\mathcal{T}$ & $S_{4}$ & $\{^{1}E,{}^{2}E\}$ & $S_{4z}=i\sigma_{z}$, $\mathcal{T}=\sigma_{x}\mathcal{K}$ &  & \tabularnewline
		\cline{1-7}
		\multirow{2}{*}{51} & \multirow{2}{*}{$p4/m$} & \multirow{2}{*}{$\Gamma$, $M$} & \multirow{2}{*}{$\{C_{4z}^{+}|00\}$, $\mathcal{P}$, $\mathcal{T}$} & \multirow{4}{*}{$C_{4h}$} & $\{^{1}E_{g},^{2}E_{g}\}$ & $C_{4z}=i\sigma_{z}$, $\mathcal{P}=\sigma_{0}$, $\mathcal{T}=\sigma_{x}\mathcal{K}$ &  & \tabularnewline
		&  &  &  &  & $\{^{1}E_{u},^{2}E_{u}\}$ & $C_{4z}=i\sigma_{z}$, $\mathcal{P}=-\sigma_{0}$, $\mathcal{T}=\sigma_{x}\mathcal{K}$ &  & \tabularnewline
		\cline{1-4} \cline{6-7}
		\multirow{2}{*}{52} & \multirow{2}{*}{$p4/n$} & \multirow{2}{*}{$\Gamma$} & \multirow{2}{*}{$\{C_{4z}^{+}|\frac{1}{2}0\}$, $\mathcal{P}$, $\mathcal{T}$} &  & $\{^{1}E_{g},{}^{2}E_{g}\}$ & $C_{4z}=i\sigma_{z}$, $\mathcal{P}=\sigma_{0}$, $\mathcal{T}=\sigma_{x}\mathcal{K}$ &  & \tabularnewline
		&  &  &  &  & $\{^{1}E_{u},{}^{2}E_{u}\}$ & $C_{4z}=i\sigma_{z}$, $\mathcal{P}=-\sigma_{0}$, $\mathcal{T}=\sigma_{x}\mathcal{K}$ &  & \tabularnewline
		\hline 
		53 & $p422$ & $\Gamma$, $M$ & $\{C_{4z}^{+}|00\}$, $\{C_{2y}|00\}$, $\mathcal{T}$ & \multirow{2}{*}{$D_{4}$} & \multirow{2}{*}{$E$} & \multirow{2}{*}{$C_{4z}=i\sigma_{z}$, $C_{2y}=-\sigma_{x}$, $\mathcal{T}=\sigma_{x}\mathcal{K}$} & \multirow{16}{*}{$ak_{+}^{2}+bk_{-}^{2}$} & \multirow{16}{*}{0}\tabularnewline
		\cline{1-4}
		54 & $p42_{1}2$ & $\Gamma$ & $\{C_{4z}^{+}|00\}$, $\{C_{2y}|\frac{1}{2}\frac{1}{2}\}$, $\mathcal{T}$ &  &  &  &  & \tabularnewline
		\cline{1-7}
		55 & $p4mm$ & $\Gamma$, $M$ & $\{C_{4z}^{+}|00\}$, $\{M_{y}|00\}$, $\mathcal{T}$ & \multirow{2}{*}{$C_{4v}$} & \multirow{2}{*}{$E$} & \multirow{2}{*}{$C_{4z}=i\sigma_{z}$, $M_{y}=\sigma_{x}$, $\mathcal{T}=\sigma_{x}\mathcal{K}$} &  & \tabularnewline
		\cline{1-4}
		56 & $p4bm$ & $\Gamma$ & $\{C_{4z}^{+}|00\}$, $\{M_{y}|\frac{1}{2}\frac{1}{2}\}$, $\mathcal{T}$ &  &  &  &  & \tabularnewline
		\cline{1-7}
		\multirow{2}{*}{61} & \multirow{2}{*}{$p4/mmm$} & \multirow{2}{*}{$\Gamma$, $M$} & \multirow{2}{*}{$\{C_{4z}^{+}|00\}$, $\{C_{2y}|00\}$, $\mathcal{P}$, $\mathcal{T}$} & \multirow{8}{*}{$D_{4h}$} & $E_{g}$ & $C_{4z}=i\sigma_{z}$, $C_{2y}=\sigma_{x}$, $\mathcal{P}=\sigma_{0}$,
		$\mathcal{T}=\sigma_{x}\mathcal{K}$ &  & \tabularnewline
		&  &  &  &  & $E_{u}$ & $C_{4z}=i\sigma_{z}$, $C_{2y}=-\sigma_{x}$, $\mathcal{P}=-\sigma_{0}$,
		$\mathcal{T}=\sigma_{x}\mathcal{K}$ &  & \tabularnewline
		\cline{1-4} \cline{6-7}
		\multirow{2}{*}{62} & \multirow{2}{*}{$p4/nbm$} & \multirow{2}{*}{$\Gamma$} & \multirow{2}{*}{$\{C_{4z}^{+}|\frac{1}{2}0\}$, $\{C_{2y}|\frac{1}{2}0\}$, $\mathcal{P}$,
		$\mathcal{T}$} &  & $E_{g}$ & $C_{4z}=i\sigma_{z}$, $C_{2y}=\sigma_{x}$, $\mathcal{P}=\sigma_{0}$,
		$\mathcal{T}=\sigma_{x}\mathcal{K}$ &  & \tabularnewline
		&  &  &  &  & $E_{u}$ & $C_{4z}=i\sigma_{z}$, $C_{2y}=-\sigma_{x}$, $\mathcal{P}=-\sigma_{0}$,
		$\mathcal{T}=\sigma_{x}\mathcal{K}$ &  & \tabularnewline
		\cline{1-4} \cline{6-7}
		\multirow{2}{*}{63} & \multirow{2}{*}{$p4/mbm$} & \multirow{2}{*}{$\Gamma$} & \multirow{2}{*}{$\{C_{4z}^{+}|00\}$, $\{C_{2y}|\frac{1}{2}\frac{1}{2}\}$, $\mathcal{P}$,
			$\mathcal{T}$} &  & $E_{g}$ & $C_{4z}=i\sigma_{z}$, $C_{2y}=\sigma_{x}$, $\mathcal{P}=\sigma_{0}$,
		$\mathcal{T}=\sigma_{x}\mathcal{K}$ &  & \tabularnewline
		&  &  &  &  & $E_{u}$ & $C_{4z}=i\sigma_{z}$, $C_{2y}=-\sigma_{x}$, $\mathcal{P}=-\sigma_{0}$,
		$\mathcal{T}=\sigma_{x}\mathcal{K}$ &  & \tabularnewline
		\cline{1-4} \cline{6-7}
		\multirow{2}{*}{64} & \multirow{2}{*}{$p4/nmm$} & \multirow{2}{*}{$\Gamma$} & \multirow{2}{*}{$\{C_{4z}^{+}|\frac{1}{2}0\}$, $\{C_{2y}|0\frac{1}{2}\}$, $\mathcal{P}$,
		$\mathcal{T}$} &  & $E_{g}$ & $C_{4z}=i\sigma_{z}$, $C_{2y}=\sigma_{x}$, $\mathcal{P}=\sigma_{0}$,
		$\mathcal{T}=\sigma_{x}\mathcal{K}$ &  & \tabularnewline
		&  &  &  &  & $E_{u}$ & $C_{4z}=i\sigma_{z}$, $C_{2y}=-\sigma_{x}$, $\mathcal{P}=-\sigma_{0}$,
		$\mathcal{T}=\sigma_{x}\mathcal{K}$ &  & \tabularnewline
		\cline{1-7}
		57 & $p\bar{4}2m$ & $\Gamma,M$ & $\{S_{4z}^{+}|00\}$, $\{C_{2y}|00\}$, $\mathcal{T}$ & \multirow{4}{*}{$D_{2d}$} & \multirow{4}{*}{$E$} & \multirow{2}{*}{$S_{4z}=i\sigma_{z}$, $C_{2y}=-\sigma_{x}$, $\mathcal{T}=\sigma_{x}\mathcal{K}$} &  & \tabularnewline
		\cline{1-4} \cline{2-4} \cline{3-4} \cline{4-4} 
		58 & $p\bar{4}2_{1}m$ & $\Gamma$ & $\{S_{4z}^{+}|00\}$, $\{C_{2y}|\frac{1}{2}\frac{1}{2}\}$, $\mathcal{T}$ &  &  &  &  & \tabularnewline
		\cline{1-4} \cline{2-4} \cline{3-4} \cline{4-4} \cline{7-7} 
		59 & $p\bar{4}m2$ & $\Gamma,M$ & $\{S_{4z}^{+}|00\}$, $\{M_{y}|00\}$, $\mathcal{T}$ &  &  & \multirow{2}{*}{$S_{4z}=i\sigma_{z}$, $M_{y}=\sigma_{x}$, $\mathcal{T}=\sigma_{x}\mathcal{K}$} &  & \tabularnewline
		\cline{1-4} \cline{2-4} \cline{3-4} \cline{4-4} 
		60 & $p\bar{4}b2$ & $\Gamma$ & $\{S_{4z}^{+}|00\}$, $\{M_{y}|\frac{1}{2}\frac{1}{2}\}$, $\mathcal{T}$ &  &  &  &  & \tabularnewline
		\hline \hline
	\end{tabular}
}
\end{table}

\

\textit{\textcolor{blue}{QNPs dictated by threefold or sixfold rotation symmetries ---}}\label{SM:sec_kdotp_QNPwosoc_C3}
It is possible to realize a QNP under threefold rotation symmetries. For example, a QNP can be obtained at point $\Gamma$ $(0,0)$ in the LG 65 ($p3$). The corresponding little co-group is $C_{3}$, with one generator $C_{3z}\equiv C_{3z}^{+}$. In the basis of IRR $\{^{1}E,~^{2}E\}$ \cite{Koster1963Properties-MP,Bradley2009Mathematical-OUP}, e.g. $\{x+iy, x-iy\}$, the matrix representation of symmetry operations are written as
\begin{equation}
C_{3z} = e^{i\frac{2\pi}{3} \sigma_{z}}, \qquad \mathcal{T}=\sigma_{x}\mathcal{K}
\end{equation}
and the symmetry constraints can be expressed as
\begin{subequations}\begin{eqnarray}
C_{3z}\mathcal{H}_{\text{eff}}(\bm{k})C_{3z}^{-1} & = & \mathcal{H}_{\text{eff}}(R_{3z}\bm{k}), \\
\mathcal{T}\mathcal{H}_{\text{eff}}(\bm{k})\mathcal{T}^{-1} & = & \mathcal{H}_{\text{eff}}(-k_{x},-k_{y}),
\end{eqnarray}\end{subequations}
where $R_{3z}$ is the threefold rotation acting on $\bm{k}$. Then, the effective Hamiltonian up to second order can be expressed as
\begin{equation}
\mathcal{H}_{\text{eff}}^{65,\ \Gamma}(\bm{k}) = \epsilon(\bm{k}) + [\alpha k_{+}^{2}\sigma_{+} + \text{H.c.}]
\end{equation}
which describes a nodal point with quadratic splitting in the $k_{x}$-$k_{y}$ plane.

One can then construct the supergroup of LG 65 to search for other possible QNPs. In Table~\ref{SM:table_C3C6}, we list the possible supergroups of LG 65 and present the corresponding effective Hamiltonian of QNPs. 

\begin{table}[!t]
	\centering
	\caption{Effective Hamilonian up to third order for the supergroups of layer groups 65 ($p3$) and 73 ($p6$) in the absence of spin-orbit coupling. LG, HSP, PG, IRR respectively stand for the layer group, the high-symmetry points (HSPs), the little co-groups of HSPs, and the irreducible representation. $f(\bm{k})$ and $g(\bm{k})$ are functions given in Eq.~\ref{SM:eq_Heff}.}\label{SM:table_C3C6}
	\scalebox{0.9}{
	\renewcommand{\arraystretch}{1.5}
	\begin{tabular}{ c|c|c|c|c|c|c|c|c }
		\hline \hline
		No. & LG & HSP & Generators & PG & IRR & Matrix Representation & $f(\bm{k})$ & $g(\bm{k})$\tabularnewline
		\hline 
		65 & $p3$ & $\Gamma$ & $\{C_{3z}^{+}|00\}$, $\mathcal{T}$ & $C_{3}$ & $\{^{1}E,^{2}E\}$ & $C_{3z}=e^{i\frac{2\pi}{3}\sigma_{z}}$, $\mathcal{T}=\sigma_{x}\mathcal{K}$ & \multirow{3}{*}{$\alpha k_{+}^{2}$} & \multirow{3}{*}{$\gamma k_{+}^{3} + \gamma^{*} k_{-}^{3}$} \tabularnewline
		\cline{1-7}
		\multirow{2}{*}{74} & \multirow{2}{*}{$p\bar{6}$} & \multirow{2}{*}{$\Gamma$} & \multirow{2}{*}{$\{C_{3z}^{+}|00\}$, $\{M_{z}|00\}$, $\mathcal{T}$} & \multirow{2}{*}{$C_{3h}$} & $\{^{1}E',^{2}E'\}$ & $C_{3z}=e^{i\frac{2\pi}{3}\sigma_{z}}$, $M_{z}=\sigma_{0}$, 		$\mathcal{T}=\sigma_{x}\mathcal{K}$ & & \tabularnewline
		&  &  &  &  & $\{^{1}E'',^{2}E''\}$ & $C_{3z}=e^{i\frac{2\pi}{3}\sigma_{z}}$, $M_{z}=-\sigma_{0}$,
		$\mathcal{T}=\sigma_{x}\mathcal{K}$ &  & \tabularnewline
		\hline 
		\multirow{2}{*}{66} & \multirow{2}{*}{$p\bar{3}$} & \multirow{2}{*}{$\Gamma$} & \multirow{2}{*}{$\{C_{3z}^{+}|00\}$, $\mathcal{P}$, $\mathcal{T}$} & \multirow{2}{*}{$C_{3i}$} & $\{^{1}E_{g},^{2}E_{g}\}$ & $C_{3z}=e^{i\frac{2\pi}{3}\sigma_{z}}$, $\mathcal{P}=\sigma_{0}$, 		$\mathcal{T}=\sigma_{x}\mathcal{K}$ & \multirow{2}{*}{$\alpha k_{+}^{2}$} & \multirow{2}{*}{0} \tabularnewline
		&  &  &  &  & $\{^{1}E_{u},^{2}E_{u}\}$ & $C_{3z}^{+}=e^{i\frac{2\pi}{3}\sigma_{z}}$, $\mathcal{P}=-\sigma_{0}$,
		$\mathcal{T}=\sigma_{x}\mathcal{K}$ &  & \tabularnewline
		
		\hline
		
		67 & $p312$ & $\Gamma$ & $\{C_{3z}^{+}|00\}$, $\{C_{2y}|00\}$, $\mathcal{T}$ & \multirow{2}{*}{$D_{3}$} & \multirow{2}{*}{$E$} & $C_{3z}=e^{i\frac{2\pi}{3}\sigma_{z}}$, $C_{2y}=-\sigma_{x}$,
		$\mathcal{T}=\sigma_{x}\mathcal{K}$ & $ak_{+}^{2}$ & $c(k_{+}^{3} + k_{-}^{3})$ \tabularnewline
		\cline{1-4} \cline{2-4} \cline{3-4} \cline{4-4} \cline{7-9} 
		68 & $p321$ & $\Gamma$ & $\{C_{3z}^{+}|00\}$, $\{C_{2x}|00\}$, $\mathcal{T}$ &  &  & $C_{3z}=e^{i\frac{2\pi}{3}\sigma_{z}}$, $C_{2x}=\sigma_{x}$,
		$\mathcal{T}=\sigma_{x}\mathcal{K}$ & $ak_{+}^{2}$ & $ic(k_{+}^{3} - k_{-}^{3})$ \tabularnewline
		
		\hline
		
		69 & $p3m1$ & $\Gamma$ & $\{C_{3z}^{+}|00\}$, $\{M_{x}|00\}$, $\mathcal{T}$ & \multirow{2}{*}{$C_{3v}$} & \multirow{2}{*}{$E$} & $C_{3z}=e^{i\frac{2\pi}{3}\sigma_{z}}$, $M_{x}=-\sigma_{x}$,
		$\mathcal{T}=\sigma_{x}\mathcal{K}$ & $ak_{+}^{2}$ & $c(k_{+}^{3} + k_{-}^{3})$ \tabularnewline
		\cline{1-4} \cline{7-9} 
		70 & $p31m$ & $\Gamma$ & $\{C_{3z}^{+}|00\}$, $\{M_{y}|00\}$, $\mathcal{T}$ &  &  & $C_{3z}=e^{i\frac{2\pi}{3}\sigma_{z}}$, $M_{y}=\sigma_{x}$,
		$\mathcal{T}=\sigma_{x}\mathcal{K}$ & $ak_{+}^{2}$ & $ic(k_{+}^{3} - k_{-}^{3})$ \tabularnewline
		
		\hline
		
		\multirow{2}{*}{71} & \multirow{2}{*}{$p\bar{3}1m$} & \multirow{2}{*}{$\Gamma$} & \multirow{2}{*}{$\{C_{3z}^{+}|00\}$, $\{C_{2y}|00\}$, $\mathcal{P}$, $\mathcal{T}$} & \multirow{4}{*}{$D_{3d}$} & $E_{g}$ & $C_{3z}=e^{i\frac{2\pi}{3}\sigma_{z}}$, $C_{2y}=\sigma_{x}$,
		$\mathcal{P}=\sigma_{0}$, $\mathcal{T}=\sigma_{x}\mathcal{K}$ & \multirow{4}{*}{$ak_{+}^{2}$} & \multirow{4}{*}{0} \tabularnewline
		&  &  &  &  & $E_{u}$ & $C_{3z}=e^{i\frac{2\pi}{3}\sigma_{z}}$, $C_{2y}=-\sigma_{x}$,
		$\mathcal{P}=-\sigma_{0}$, $\mathcal{T}=\sigma_{x}\mathcal{K}$ &  & \tabularnewline
		\cline{1-4}\cline{6-7}
		\multirow{2}{*}{72} & \multirow{2}{*}{$p\bar{3}m1$} & \multirow{2}{*}{$\Gamma$} & \multirow{2}{*}{$\{C_{3z}^{+}|00\}$, $\{C_{2x}|00\}$, $\mathcal{P}$, $\mathcal{T}$} &  & $E_{g}$ & $C_{3z}=e^{i\frac{2\pi}{3}\sigma_{z}}$, $C_{2x}=-\sigma_{x}$, $\mathcal{P}=\sigma_{0}$, $\mathcal{T}=\sigma_{x}\mathcal{K}$ &  & \tabularnewline
		&  &  &  &  & $E_{u}$ & $C_{3z}=e^{i\frac{2\pi}{3}\sigma_{z}}$, $C_{2x}=\sigma_{x}$,
		$\mathcal{P}=-\sigma_{0}$, $\mathcal{T}=\sigma_{x}\mathcal{K}$ &  & \tabularnewline
		
		\hline
		
		\multirow{2}{*}{78} & \multirow{2}{*}{$p\bar{6}m2$} & \multirow{2}{*}{$\Gamma$} & \multirow{2}{*}{$\{C_{3z}^{+}|00\}$, $\{M_{z}|00\}$, $\{M_{x}|00\}$, $\mathcal{T}$} & \multirow{4}{*}{$D_{3h}$} & $E'$ & $C_{3z}=e^{i\frac{2\pi}{3}\sigma_{z}}$, $M_{z}=\sigma_{0}$,
		$M_{x}=-\sigma_{x}$, $\mathcal{T}=\sigma_{x}\mathcal{K}$ & \multirow{2}{*}{$ak_{+}^{2}$} & \multirow{2}{*}{$c(k_{+}^{3} +  k_{-}^{3})$} \tabularnewline
		&  &  &  &  & $E''$ & $C_{3z}=e^{i\frac{2\pi}{3}\sigma_{z}}$, $M_{z}=-\sigma_{0}$,
		$M_{x}=-\sigma_{x}$, $D(\mathcal{T})=\sigma_{x}\mathcal{K}$ &  & \tabularnewline
		\cline{1-4} \cline{6-9} 
		\multirow{2}{*}{79} & \multirow{2}{*}{$p\bar{6}2m$} & \multirow{2}{*}{$\Gamma$} & \multirow{2}{*}{$\{C_{3z}^{+}|00\}$, $\{M_{z}|00\}$, $\{C_{2x}|00\}$, $\mathcal{T}$} &  & $E'$ & $C_{3z}=e^{i\frac{2\pi}{3}\sigma_{z}}$, $M_{z}=\sigma_{0}$,
		$C_{2x}=\sigma_{x}$, $\mathcal{T}=\sigma_{x}\mathcal{K}$ & \multirow{2}{*}{$ak_{+}^{2}$} & \multirow{2}{*}{$ic(k_{+}^{3} - k_{-}^{3})$} \tabularnewline
		&  &  &  &  & $E''$ & $C_{3z}=e^{i\frac{2\pi}{3}\sigma_{z}}$, $M_{z}=-\sigma_{0}$,
		$C_{2x}=-\sigma_{x}$, $\mathcal{T}=\sigma_{x}\mathcal{K}$ &  & \tabularnewline
		\hline \hline
        \multirow{2}{*}{73} & \multirow{2}{*}{$p6$} & \multirow{2}{*}{$\Gamma$} & \multirow{2}{*}{$\{C_{6z}^{+}|00\}$, $\mathcal{T}$} & \multirow{2}{*}{$C_{6}$} & $\{^{1}E_{1},{}^{2}E_{1}\}$ & $C_{6z}=e^{i\frac{\pi}{3}\sigma_{z}}$, $\mathcal{T}=\sigma_{x}\mathcal{K}$ & \multirow{6}{*}{$\alpha k_{+}^{2}$} & \multirow{6}{*}{0}\tabularnewline
        &  &  &  &  & $\{^{1}E_{2},{}^{2}E_{2}\}$ & $C_{6z}=-e^{i\frac{\pi}{3}\sigma_{z}}$, $\mathcal{T}=\sigma_{x}\mathcal{K}$ &  & \tabularnewline
        \cline{1-7} \cline{2-7} \cline{3-7} \cline{4-7} \cline{5-7} \cline{6-7} \cline{7-7} 
        \multirow{4}{*}{75} & \multirow{4}{*}{$p6/m$} & \multirow{4}{*}{$\Gamma$} & \multirow{4}{*}{$\{C_{6z}^{+}|00\}$, $\mathcal{P}$, $\mathcal{T}$} & \multirow{4}{*}{$C_{6h}$} & $\{^{1}E_{1g},{}^{2}E_{1g}\}$ & $C_{6z}=e^{i\frac{\pi}{3}\sigma_{z}}$, $\mathcal{P}=\sigma_{0}$,
        $\mathcal{T}=\sigma_{x}\mathcal{K}$ &  & \tabularnewline
        &  &  &  &  & $\{^{1}E_{1u},{}^{2}E_{1u}\}$ & $C_{6z}=e^{i\frac{\pi}{3}\sigma_{z}}$, $\mathcal{P}=-\sigma_{0}$,
        $\mathcal{T}=\sigma_{x}\mathcal{K}$ &  & \tabularnewline
        &  &  &  &  & $\{^{1}E_{2g},{}^{2}E_{2g}\}$ & $C_{6z}=-e^{i\frac{\pi}{3}\sigma_{z}}$, $\mathcal{P}=\sigma_{0}$,
        $\mathcal{T}=\sigma_{x}\mathcal{K}$ &  & \tabularnewline
        &  &  &  &  & $\{^{1}E_{2u},{}^{2}E_{2u}\}$ & $C_{6z}=-e^{i\frac{\pi}{3}\sigma_{z}}$, $\mathcal{P}=-\sigma_{0}$,
        $\mathcal{T}=\sigma_{x}\mathcal{K}$ &  & \tabularnewline
        \hline 
        \multirow{2}{*}{76} & \multirow{2}{*}{$p622$} & \multirow{2}{*}{$\Gamma$} & \multirow{2}{*}{$\{C_{6z}^{+}|00\}$, $\{C_{2x}|00\}$, $\mathcal{T}$} & \multirow{2}{*}{$D_{6}$} & $E_{1}$ & $C_{6z}=e^{i\frac{\pi}{3}\sigma_{z}}$, $C_{2x}=\sigma_{x}$,
        $\mathcal{T}=\sigma_{x}\mathcal{K}$ & \multirow{8}{*}{$ak_{+}^{2}$} & \multirow{8}{*}{0}\tabularnewline
        &  &  &  &  & $E_{2}$ & $C_{6z}=-e^{i\frac{\pi}{3}\sigma_{z}}$, $C_{2x}=\sigma_{x}$,
        $\mathcal{T}=\sigma_{x}\mathcal{K}$ &  & \tabularnewline
        \cline{1-7} \cline{2-7} \cline{3-7} \cline{4-7} \cline{5-7} \cline{6-7} \cline{7-7} 
        \multirow{2}{*}{77} & \multirow{2}{*}{$p6mm$} & \multirow{2}{*}{$\Gamma$} & \multirow{2}{*}{$\{C_{6z}^{+}|00\}$, $\{M_{x}|00\}$, $\mathcal{T}$} & \multirow{2}{*}{$C_{6v}$} & $E_{1}$ & $C_{6z}=e^{i\frac{\pi}{3}\sigma_{z}}$, $M_{x}=-\sigma_{x}$,
        $\mathcal{T}=\sigma_{x}\mathcal{K}$ &  & \tabularnewline
        &  &  &  &  & $E_{2}$ & $C_{6z}=-e^{i\frac{\pi}{3}\sigma_{z}}$, $M_{x}=\sigma_{x}$,
        $\mathcal{T}=\sigma_{x}\mathcal{K}$ &  & \tabularnewline
        \cline{1-7} \cline{2-7} \cline{3-7} \cline{4-7} \cline{5-7} \cline{6-7} \cline{7-7} 
        \multirow{4}{*}{80} & \multirow{4}{*}{$p6/mmm$} & \multirow{4}{*}{$\Gamma$} & \multirow{4}{*}{$\{C_{6z}^{+}|00\}$, $\{M_{x}|00\}$, $\mathcal{P}$, $\mathcal{T}$} & \multirow{4}{*}{$D_{6h}$} & $E_{1g}$ & $C_{6z}=e^{i\frac{\pi}{3}\sigma_{z}}$, $\mathcal{P}=\sigma_{0}$,
        $M_{x}=-\sigma_{x}$, $\mathcal{T}=\sigma_{x}\mathcal{K}$ &  & \tabularnewline
        &  &  &  &  & $E_{1u}$ & $C_{6z}=e^{i\frac{\pi}{3}\sigma_{z}}$, $\mathcal{P}=-\sigma_{0}$,
        $M_{x}=-\sigma_{x}$, $\mathcal{T}=\sigma_{x}\mathcal{K}$ &  & \tabularnewline
        &  &  &  &  & $E_{2g}$ & $C_{6z}=-e^{i\frac{\pi}{3}\sigma_{z}}$, $\mathcal{P}=\sigma_{0}$,
        $M_{x}=\sigma_{x}$, $\mathcal{T}=\sigma_{x}\mathcal{K}$ &  & \tabularnewline
        &  &  &  &  & $E_{2u}$ & $C_{6z}=-e^{i\frac{\pi}{3}\sigma_{z}}$, $\mathcal{P}=-\sigma_{0}$,
        $M_{x}=\sigma_{x}$, $\mathcal{T}=\sigma_{x}\mathcal{K}$ &  & \tabularnewline
        \hline \hline
	\end{tabular}
}
\end{table}

It is noted that the LG 73 ($p6$) is also a supergroup of LG 65, but we list it and its supergroups in Table~\ref{SM:table_C3C6} separately. For LG 73, the QNP at $\Gamma$ $(0,0)$ is guaranteed by the IRR $\{^{1}E_{1},~^{2}E_{1}\}$ or $\{^{1}E_{2},~^{2}E_{2}\}$ of little co-group $C_{6}$ \cite{Koster1963Properties-MP,Bradley2009Mathematical-OUP}. The matrix representations of $C_{6z} \equiv C_{6z}^{+}$ under the two IRRs differ by only negative sign, say, $\pm e^{i\frac{2\pi}{3}\sigma_{z}}$. After including time-reversal $\mathcal{T}$ whose matrix representation is given by $\mathcal{T}=\sigma_{x}\mathcal{K}$ and under the symmetry constraints ($R_{6z}$ is the sixfold rotation acting on $\bm{k}$)
\begin{subequations}
\begin{eqnarray}
	C_{6z}\mathcal{H}_{\text{eff}}(\bm{k})C_{6z}^{-1} & = & \mathcal{H}_{\text{eff}}(R_{6z}\bm{k}), \\
	\mathcal{T}\mathcal{H}_{\text{eff}}(\bm{k})\mathcal{T}^{-1} & = & \mathcal{H}_{\text{eff}}(-k_{x},-k_{y}),
\end{eqnarray}
\end{subequations}
the effective Hamiltonian up to second order can be expressed as
\begin{equation}
\mathcal{H}_{\text{eff}}^{73,\ \Gamma}(\bm{k}) = \epsilon(\bm{k}) + [\alpha k_{+}^{2}\sigma_{+} + \text{H.c.}]
\end{equation}
which describes a QNP with quadratic splitting in the $k_{x}$-$k_{y}$ plane.

\subsection{Cubic nodal points in the presence of spin-orbit coupling}\label{SM:sec_kdotp_QNPwsoc}

\textit{\textcolor{blue}{CNPs dictated by threefold or sixfold rotation symmetries ---}}
In the presence of SOC, cubic nodal points (CNPs) can be realized, dictated by threefold or sixfold rotation symmetries. Taking LG 65 ($p3$) as an example, a CNP appears at $\Gamma$ $(0,0)$, whose little co-group is $C_{3}$. The generator includes $C_{3z}\equiv C_{3z}^{+}$ which satisfies
\begin{equation}
C_{3z}^{3} = -1
\end{equation}
and thus has the eigenvalues $c_{3z} = e^{\pm i\pi/3}, -1$. Considering time-reversal $\mathcal{T}$, the CNP corresponds to the IRR $\{\bar{E}, \bar{E}\}$. Then, one can choose the basis as $\{|c_{3z}=-1\rangle, \mathcal{T}|-1\rangle\}$ \cite{Koster1963Properties-MP,Bradley2009Mathematical-OUP}, in which the matrix representations of symmetry operations are expressed as
\begin{equation}
C_{3z} = -\sigma_{0} \qquad
\mathcal{T} = -i\sigma_{y}\mathcal{K}
\end{equation}
The symmetry constraints on Hamiltonian are
\begin{equation*}
\begin{aligned}
C_{3z}\mathcal{H}(k_{x},k_{y})C_{3z}^{-1} &= \mathcal{H}(R_{3z}\bm{k}) \\
\mathcal{T}\mathcal{H}(k_{x},k_{y})\mathcal{T}^{-1} &= \mathcal{H}(-k_{x},-k_{y}) \\
\end{aligned}
\end{equation*}
A direct calculation up to third order gives
\begin{equation}
\mathcal{H}_{\text{eff}}^{65,\ \Gamma} = \epsilon(\bm{k})
+ [(\alpha k_{+}^3 + \beta k_{-}^3)\sigma_{+} + \text{H.c.}] + (\gamma k_{+}^3 + \gamma^{*} k_{-}^3) \sigma_{z} 
\end{equation}
where $\epsilon(\bm{k}) = \omega_{0} + \omega_{1}(k_{x}^{2} + k_{y}^{2})$, and $\alpha$, $\beta$ and $\gamma$ are complex parameters. This describes a CNP with cubic splitting in the $k_{x}$-$k_{y}$ plane.

Under SOC, the existence of doubly degenerate point requires the broken space-time inversion symmetry $\mathcal{PT}$; otherwise, it would become fourfold degenerate. Through constructing the noncentrosymmetric supergroup of LG 65, we list other possible CNPs and the corresponding effective Hamiltonian is given in Table~\ref{SM:table_wsoc}. It is noted that although the LGs 74,78, and 79 do not have inversion symmetry, their CNPs at $\Gamma$ is actually not an isolated point but resides on three unremovable nodal lines, as discussed later.

\begin{table}[!t]
	\centering
	\caption{Effective Hamilonian up to fourth order in the presence of spin-orbit coupling. LG, HSP, PG, IRR stand for the layer group, the high-symmetry points (HSPs), the little co-groups of HSPs, and the irreducible representation, respectively. $f(\bm{k})$ and $g(\bm{k})$ are functions given in Eq.~\ref{SM:eq_Heff}.}\label{SM:table_wsoc}
	\scalebox{1.0}{
		\renewcommand{\arraystretch}{1.8}
		\begin{tabular}{ c|c|c|c|c|c|c|c|c }
			\hline \hline
			No. & LG & HSP & Generators & PG & IRR & Matrix Representation & $f(\bm{k})$ & $g(\bm{k})$\tabularnewline
			\hline 
			65 & $p3$ & $\Gamma$ & $\{C_{3z}^{+}|00\}$, $\mathcal{T}$ & $C_{3}$ & $\{\bar{E},\bar{E}\}$ & $C_{3z}=-\sigma_{0}$, $\mathcal{T}=i\sigma_{y}\mathcal{K}$ & \multirow{1}{*}{$\alpha k_{+}^{3}+\beta k_{-}^{3}$} & $\gamma k_{+}^{3}+\gamma^{*}k_{-}^{3}$\tabularnewline
			\hline 
			67 & $p312$ & $\Gamma$ & $\{C_{3z}^{+}|00\}$, $\{C_{2y}|00\}$, $\mathcal{T}$ & \multirow{2}{*}{$D_{3}$} & \multirow{2}{*}{$\{^{1}\bar{E},{}^{2}\bar{E}\}$} & $C_{3z}=-\sigma_{0}$, $C_{2y}=i\sigma_{z}$, $\mathcal{T}=i\sigma_{y}\mathcal{K}$ & $\alpha(k_{+}^{3}+k_{-}^{3})$ & $ic(k_{+}^{3}-k_{-}^{3})$\tabularnewline
			\cline{1-4} \cline{2-4} \cline{3-4} \cline{4-4} \cline{7-9} \cline{8-9} \cline{9-9} 
			68 & $p321$ & $\Gamma$ & $\{C_{3z}^{+}|00\}$, $\{C_{2x}|00\}$, $\mathcal{T}$ &  &  & $C_{3z}=-\sigma_{0}$, $C_{2x}=i\sigma_{z}$, $\mathcal{T}=i\sigma_{y}\mathcal{K}$ & $i\alpha(k_{+}^{3}-k_{-}^{3})$ & $c(k_{+}^{3}+k_{-}^{3})$\tabularnewline
			\hline 
			69 & $p3m1$ & $\Gamma$ & $\{C_{3z}^{+}|00\}$, $\{M_{x}|00\}$, $\mathcal{T}$ & \multirow{2}{*}{$C_{3v}$} & \multirow{2}{*}{$\{^{1}\bar{E},{}^{2}\bar{E}\}$} & $C_{3z}=-\sigma_{0}$, $M_{x}=i\sigma_{z}$, $\mathcal{T}=i\sigma_{y}\mathcal{K}$ & $\alpha(k_{+}^{3}+k_{-}^{3})$ & $ic(k_{+}^{3}-k_{-}^{3})$\tabularnewline
			\cline{1-4} \cline{2-4} \cline{3-4} \cline{4-4} \cline{7-9} \cline{8-9} \cline{9-9} 
			70 & $p31m$ & $\Gamma$ & $\{C_{3z}^{+}|00\}$, $\{M_{y}|00\}$, $\mathcal{T}$ &  &  & $C_{3z}=-\sigma_{0}$, $M_{y}=i\sigma_{z}$, $\mathcal{T}=i\sigma_{y}\mathcal{K}$ & $i\alpha(k_{+}^{3}-k_{-}^{3})$ & $c(k_{+}^{3}+k_{-}^{3})$\tabularnewline
			\hline 
			\hline 
			73 & $p6$ & $\Gamma$ & $\{C_{6z}^{+}|00\}$, $\mathcal{T}$ & $C_{6}$ & $\{^{1}\bar{E}_{1},{}^{2}\bar{E}_{1}\}$ & $C_{6z}=i\sigma_{z}$, $\mathcal{T}=i\sigma_{y}\mathcal{K}$ & $\alpha k_{+}^{3}+\beta k_{-}^{3}$ & 0\tabularnewline
			\hline 
			76 & $p622$ & $\Gamma$ & $\{C_{6z}^{+}|00\}$, $\{C_{2x}|00\}$, $\mathcal{T}$ & $D_{6}$ & $\bar{E}_{3}$ & $C_{6z}=i\sigma_{z}$, $C_{2x}=i\sigma_{x}$, $\mathcal{T}=i\sigma_{y}\mathcal{K}$ & $ak_{+}^{3}+bk_{-}^{3}$ & 0\tabularnewline
			\hline
			77 & $p6mm$ & $\Gamma$ & $\{C_{6z}^{+}|00\}$, $\{M_{y}|00\}$, $\mathcal{T}$ & $C_{6v}$ & $\bar{E}_{3}$ & $C_{6z}=i\sigma_{z}$, $M_{x}=i\sigma_{x}$, $\mathcal{T}=i\sigma_{y}\mathcal{K}$ & $iak_{+}^{3}+ibk_{-}^{3}$ & 0\tabularnewline
			\hline \hline
		\end{tabular}
	}
\end{table}

Since LG 73 ($p6$) is a supergroup of LG 65 ($p3$), the corresponding possible CNPs in the supergroups of LG 73 are also listed in Table~\ref{SM:table_wsoc}. Here, we take $p6$ as an example to demonstrate the effects of sixfold rotation symmetry on CNPs. The little co-group at $\Gamma$ is $C_{6}$ with the generator of $C_{6z} \equiv C_{6z}^{+}$. Since
\begin{equation}
C_{6z}^{6} = -1
\end{equation}
the eigenvalues of $C_{6z}$ are given as $c_{6z} = e^{\pm i\pi/6}, e^{\pm i5\pi/6}, \pm i$. Considering time-reversal $\mathcal{T}$, the CNP corresponds to the IRR $\{^{1}\bar{E}_{1},~^{2}\bar{E}_{1}\}$. By choosing the basis $\{|c_{6z} = +i\rangle, \mathcal{T}|+i\rangle\}$, the matrix representation of $C_{6z}$ and $\mathcal{T}$ can be expressed as \cite{Koster1963Properties-MP,Bradley2009Mathematical-OUP}
\begin{equation*}
C_{6z} = i\sigma_{z} \qquad
\mathcal{T} = i\sigma_{y}\mathcal{K}.
\end{equation*}
Under the symmetry constraints
\begin{equation*}
\begin{aligned}
C_{6z}\mathcal{H}(k_{x},k_{y})C_{6z}^{-1} &= \mathcal{H}(R_{6z}\bm{k}), \\
\mathcal{T}\mathcal{H}(k_{x},k_{y})\mathcal{T}^{-1} &= \mathcal{H}(-k_{x},-k_{y}),
\end{aligned}
\end{equation*}
the effective Hamiltonian up to third order gives
\begin{equation}
\mathcal{H}_{\text{eff}}^{73,\ \Gamma} = \epsilon(\bm{k})
+ [(\alpha k_{+}^3 + \beta k_{-}^3)\sigma_{+} + H.c.]
\end{equation}
which describes a CNP with cubic splitting in the $k_{x}$-$k_{y}$ plane.

\begin{figure}[!t]
	\centering
	\includegraphics[width=0.7\textwidth]{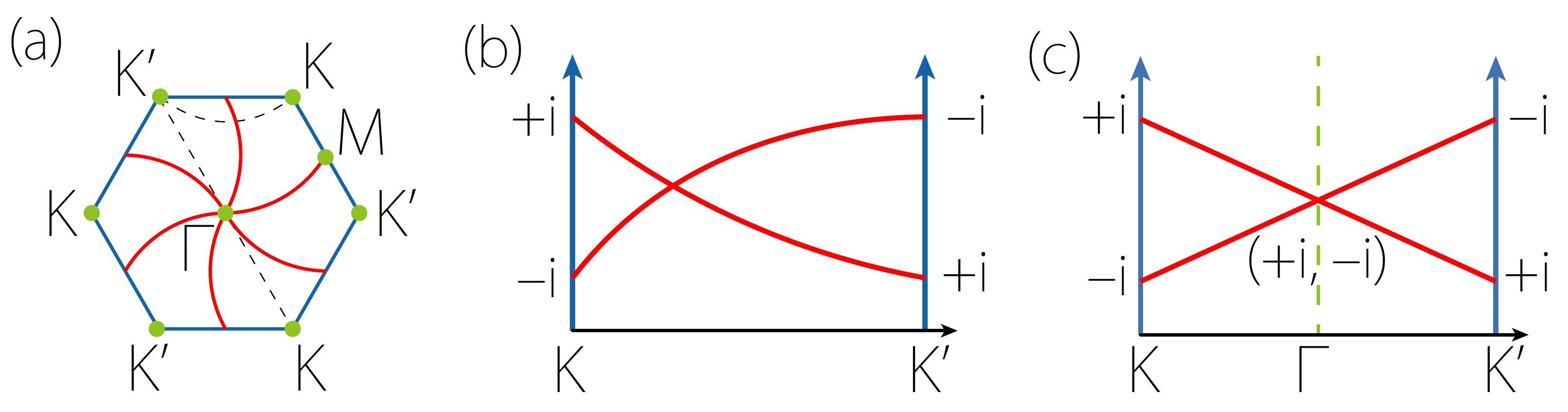}
	\caption{(a) Schematic pattern showing the inevitable nodal lines (as indicated by the red lines) in LG 74 under SOC. A generic path connecting $K$ and $K'$, as well as the $K$-$\Gamma$-$K'$ path are indicated by the dashed lines. (b) Schematic figure presenting inevitable band crossing along the generic path from $K$ to $K'$ shown in (a). The labels indicate the $M_z$ eigenvalue. The partner switching between two states with opposite $M_z$ eigenvalues leads to a twofold crossing point. (c) The inevitable band crossing point is fixed at $\Gamma$ due to the time-reversal-symmetry-induced Kramers degeneracy.}
	\label{SM:fig_LG74}
\end{figure}


\textit{\textcolor{blue}{Inevitable nodal lines ---}}\label{SM:sec_kdotp_QNPwsoc_NL}
Here, we give the symmetry analysis on the inevitable spin-orbit nodal line in LGs 74, 78, and 79. First, we consider a 2D non-magnetic hexagonal lattice with non-negligible SOC. The time reversal symmetry $\mathcal{T}$ is present, and $\mathcal{T}^{2}=-1$. To enable nodal lines, we need the horizontal mirror symmetry $M_{z}$ whose invariant plane is the entire 2D Brillouin zone (BZ). Since $[M_{z}, \mathcal{H}(\bm{k})] = 0$, the energy eigenstate can be chosen the eigenstates of $M_{z}$, which we denote as $|m_{z}\rangle$ with the $M_{z}$ eigenvalue $m_{z} \in \{\pm i\}$.

As shown in Fig.~\ref{SM:fig_LG74}(b), for a state $|m_{z} \rangle$ with the mirror eigenvalue $m_{z}$ at $K$, $\mathcal{T}|m_{z}\rangle$ at $K'$ must have the opposite eigenvalue $-m_{z}$. Then there must be a switch of partner when going from $K$ point to $K'$ point along an arbitrary path, during which two bands must cross each other. At the time-reversal-invariant-momenta (TRIMs), each Kramers pair $|m_{z} \rangle$ and $\mathcal{T}|m_{z} \rangle$ must have opposite $m_{z}$. Hence, if the path pass through a TRIM point, such as $\Gamma$ and $M$, the crossing point would be pinned at the TRIM point, as shown in Fig.~\ref{SM:fig_LG74}(c). As a result, if no extra symmetries are present, it would form three inevitable nodal lines as displayed in Fig.~\ref{SM:fig_LG74}(a). This is just the case of LG 74.

If extra vertical mirrors are present, such as in LGs 78 and 79, the spin-orbit nodal lines would be fixed on the mirror-invariant lines.

\section{Lattice model for layer group 65 without spin-orbit coupling}

\begin{figure}[t]
	\centering
	\includegraphics[width=0.8\textwidth]{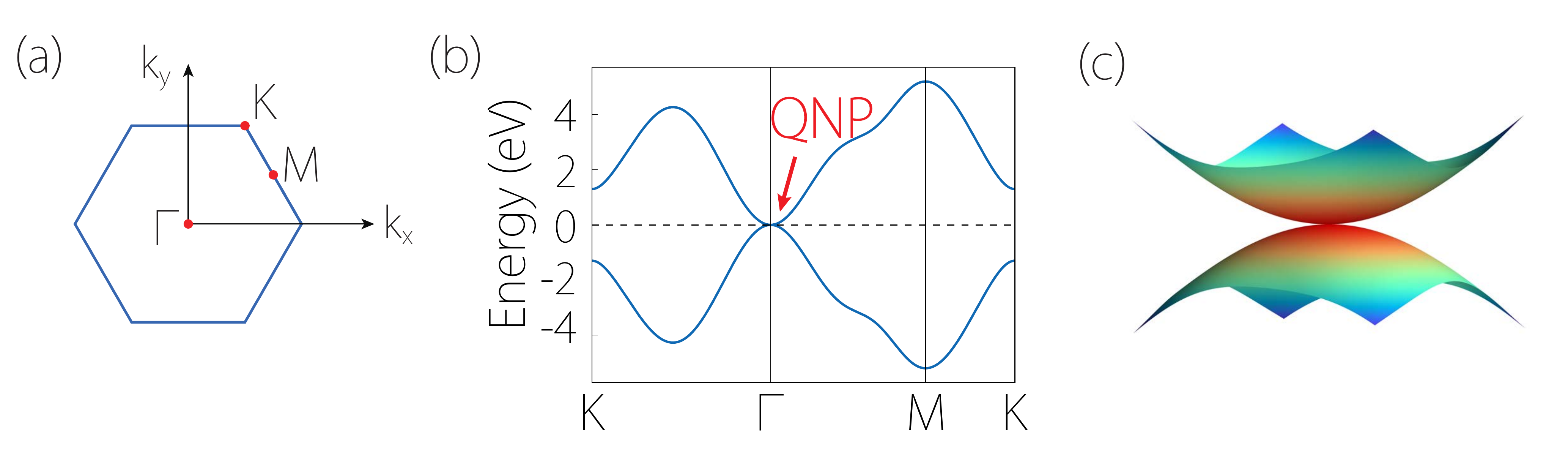}
	\caption{(a) Brillouin zone for LG 65. (b) Band structure for the spinless lattice model in Eq.~\eqref{SM:eq_Latt65}. A QNP is located at $\Gamma$ point (as indicated by the red arrow). (c) The quadratic energy dispersion around the QNP. In the figure, the model parameters are set to $t_0 = 0.0$, $t_1 = 0.0$, $t_2 = -1.0$, $t_3 = 0.6$, $t_4 = -0.5$, $t_5 = 0.0$, $t_6 = 0.5$, $t_7 = 0.6$, and $t_8 = 0.0$.}
	\label{sfig:LG65_wsoc}
\end{figure}

In this section, we demonstrate the existence of QNPs in LG 65 through the minimal lattice model. We choose a triangular lattice with a single site $(0,0,0)$ ($1a$ Wyckoff position) to construct the lattice model. Two orbitals with the IRRs $\{^{1}E,~^{2}E\}$ of the $C_{3}$ group are placed on each site. Under this basis, the generators of LG 65 can be represented as
\begin{subequations}\begin{eqnarray}
		C_{3z} &=& e^{i\frac{2\pi}{3} \sigma_{z}} \otimes (k_{x} \rightarrow \cos{\frac{2\pi}{3}} k_{x} - \sin{\frac{2\pi}{3}} k_{y}, k_{y} \rightarrow \sin{\frac{2\pi}{3}} k_{x} + \cos{\frac{2\pi}{3}} k_{y}), \\
		\mathcal{T} &=& \sigma_{x}\mathcal{K} \otimes (\bm{k} \rightarrow -\bm{k}),
\end{eqnarray}\end{subequations}
where $\sigma_i$ are Pauli matrices acting on the orbital space. Then, the symmetry-allowed lattice Hamiltonian can be express as (up to the second-neighbor hopping)
\begin{equation}\label{SM:eq_Latt65}
	\begin{aligned}
		H_{\text{latt}}^{\text{LG65}} = &t_{0}\sigma_{0} + t_{1}\sigma_{0}(\cos k_{x} + \cos(-\frac{k_{x}}{2} + \frac{\sqrt{3}k_{y}}{2}) + \cos(-\frac{k_{x}}{2} - \frac{\sqrt{3}k_{y}}{2})) \\
		&+ [(t_{2} + it_{3})\sigma_{+}(\cos k_{x} + e^{-i\frac{2\pi}{3}} \cos(-\frac{k_{x}}{2} + \frac{\sqrt{3}k_{y}}{2}) + e^{i\frac{2\pi}{3}} \cos(-\frac{k_{x}}{2} - \frac{\sqrt{3}k_{y}}{2})) + \text{H.c.}] \\
		&+ t_{4}\sigma_{z}(\sin k_{x} + \sin(-\frac{k_{x}}{2} + \frac{\sqrt{3}k_{y}}{2}) + \sin(-\frac{k_{x}}{2} - \frac{\sqrt{3}k_{y}}{2})) \\
		&+ t_{5}\sigma_{0}(\cos(-\sqrt{3}k_{y}) + \cos(\frac{3k_{x}}{2} + \frac{\sqrt{3}k_{y}}{2}) + \cos(-\frac{3k_{x}}{2} + \frac{\sqrt{3}k_{y}}{2})) \\
		&+ [(t_{6} + it_{7})\sigma_{+}(\cos(-\sqrt{3}k_{y}) + e^{-i2\pi/3}\cos(\frac{3k_{x}}{2} + \frac{\sqrt{3}k_{y}}{2}) + e^{i2\pi/3}\cos(-\frac{3k_{x}}{2} + \frac{\sqrt{3}k_{y}}{2})) + \text{H.c.}] \\
		&+ t_{8}\sigma_{z}(\sin(-\sqrt{3}k_{y}) + \sin(\frac{3k_{x}}{2} + \frac{\sqrt{3}k_{y}}{2}) + \sin(-\frac{3k_{x}}{2} + \frac{\sqrt{3}k_{y}}{2})) \\
	\end{aligned}
\end{equation}
where $\sigma_{\pm} = \frac{1}{2}(\sigma_{x} \pm i\sigma_{y})$, and $t$'s are real parameters. Through expanding this Hamiltonian at $\Gamma$, one recovers the QNP effective model shown in the main text.

Using the constructed lattice model, we plot a typical band structure shown in Fig~\ref{sfig:LG65_wsoc}. A QNP can be clearly observed at $\Gamma$. Interestingly, except for the QNP, there are no other band crossings near the Fermi energy, which indicates the existence of a single QNP in the BZ.

\section{Materials candidates}

\begin{figure}[t]
	\centering
	\includegraphics[width=0.65\textwidth]{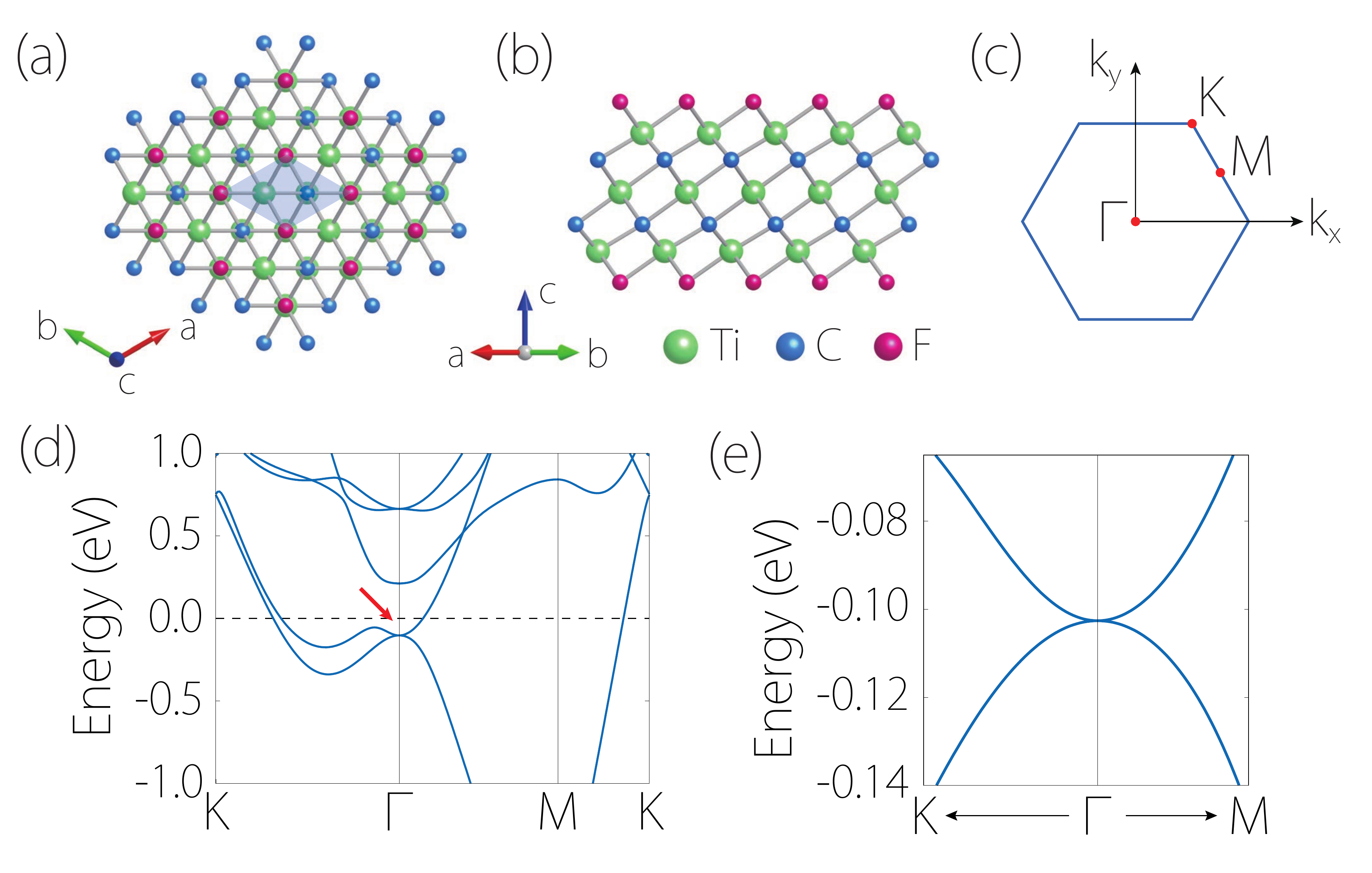}
	\caption{(a) Top and (b) side view of the crystal structure of monolayer Ti$_3$C$_2$F$_2$. (c) 2D Brillouin zone. (d) Calculated band structure for Ti$_3$C$_2$F$_2$ monolayer (without SOC). The red arrows indicate the QNP at $\Gamma$. The enlarged view of the band dispersion around the QNP is plotted in (e).}
	\label{sfig:Ti3C2F2}
\end{figure}

The symmetry conditions provide useful guides for the materials search. Here, we present a few 2D material candidates which possess higher-order nodal points. 
The first example is the 2D QNP material Ti$_3$C$_2$F$_2$ whose crystal structure has the symmetry of LG 72 ($p\bar{3}m1$) as illustrated in Fig.~\ref{sfig:Ti3C2F2}(a) and (b). Ti$_3$C$_2$T$_2$ was the first 2D MXene synthesized in 2011 \cite{Naguib2011Two-AM}. The calculated band structure of Ti$_3$C$_2$T$_2$ in the absence of SOC is displayed in Fig.~\ref{sfig:Ti3C2F2}(d), which exhibits a QNP near the Fermi energy at $\Gamma$ point. The zoom-in image for the quadratic dispersion is given in Fig.~\ref{sfig:Ti3C2F2}(e). The corresponding IRR for the QNP is $E_g$ of $D_{3d}$ point group, in agreement with our symmetry analysis.

\begin{figure}[t]
	\centering
	\includegraphics[width=0.65\textwidth]{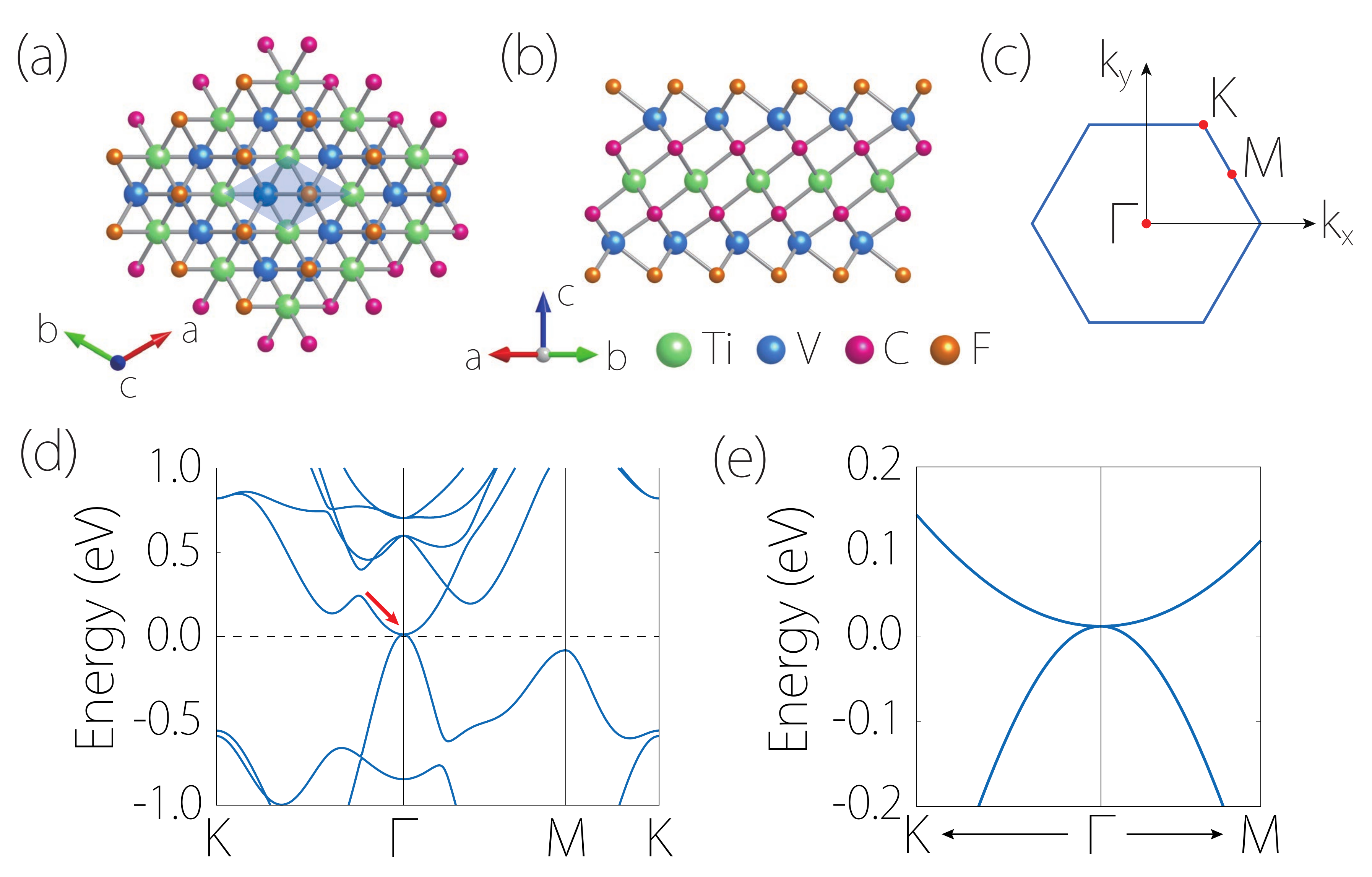}
	\caption{(a) Top and (b) side view of the crystal structure of monolayer V$_2$TiC$_2$F$_2$. (c) shows the 2D Brillouin zone. (d) Calculated band structure for V$_2$TiC$_2$F$_2$ (without SOC). The red arrows indicate the QNP at $\Gamma$. (e) gives the enlarged view of the band dispersion around the QNP in (d).}
	\label{sfig:V2TiC2F2}
\end{figure}

The second candidates is the QNP material V$_2$TiC$_2$F$_2$ with the LG 72 ($p\bar{3}m1$). V$_2$TiC$_2$F$_2$ belongs to the family of functionalized double transition metal carbide MXenes ($M'_2 M''$C$_2$$X_2$ ($M'$ = V, Nb, Ta; $M''$ = Ti, Zr, Hf; $X$ = F, Cl, Br, I, O, H, or OH), which was predicted to be a topological insulator under SOC \cite{Huang2020Large-PRB}. However, the SOC-induced bandgap for V$_2$TiC$_2$F$_2$ is only 39 meV (248 meV using HSE06), indicating that its SOC strength is quite small \cite{Huang2020Large-PRB}. From the electronic band structure (without SOC) plotted in Fig.~\ref{sfig:V2TiC2F2}(d), one clearly observes a nodal point with quadratic dispersion near the Fermi energy at $\Gamma$ point. We calculate the IRR for the QNP, which gives $E_g$ of the $D_{3d}$ point group. This is consistent with our analysis.

\begin{figure}[t]
	\centering
	\includegraphics[width=0.65\textwidth]{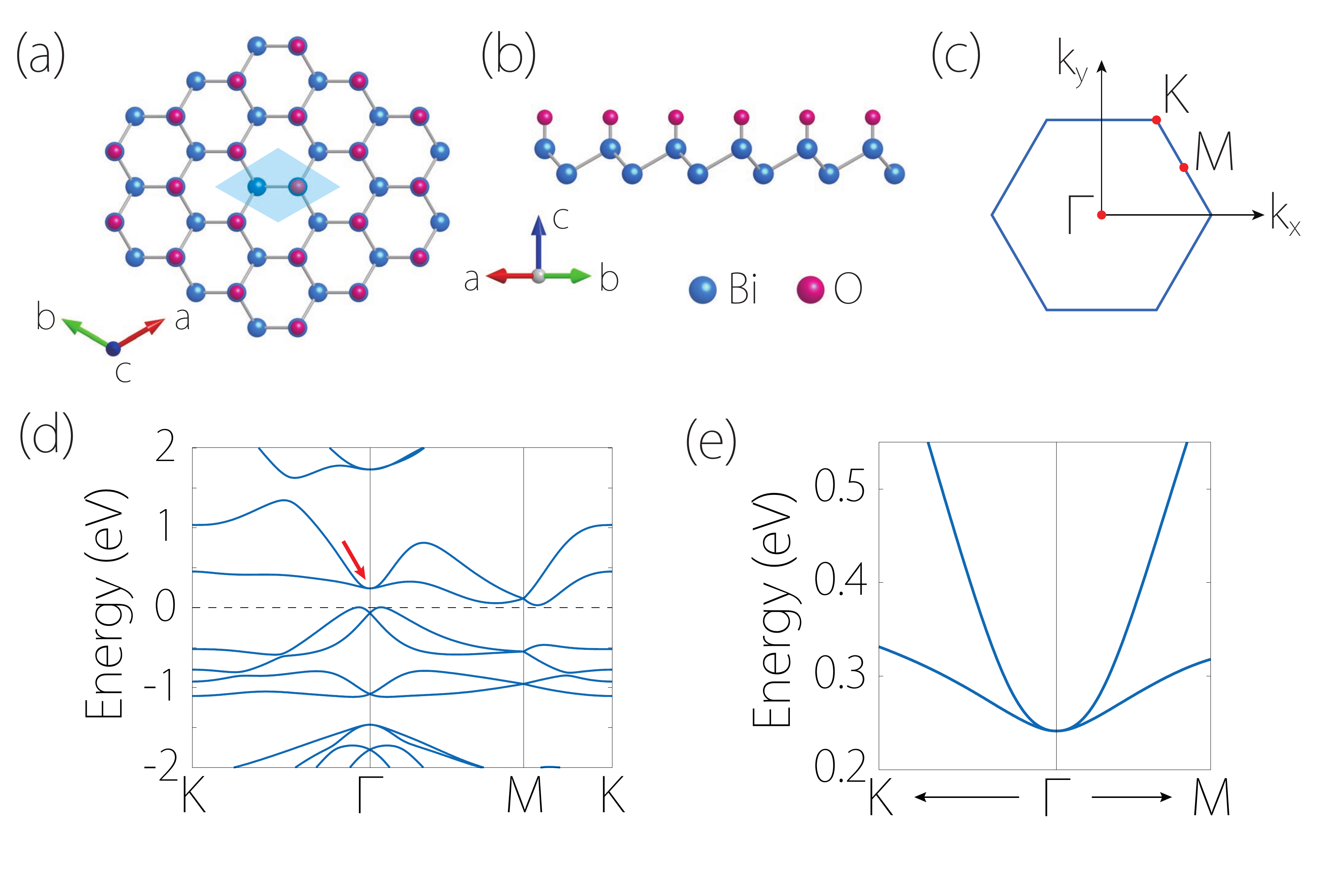}
	\caption{(a) Top and (b) side view of the crystal structure of monolayer Bi$_2$O. (c) shows the 2D Brillouin zone. (d) Calculated band structure for Bi$_2$O in the presence of SOC. The red arrows indicate the CNP at $\Gamma$. (e) gives the enlarged view of the band dispersion around the CNP in (d).}
	\label{sfig:Bi2O}
\end{figure}

The third candidate is the Bi$_2$O monolayer which hosts a CNP when SOC is considered. It was theoretically predicted through half-oxidizing the bismuthene, and exhibits the Rashba effect due to the lacking of inversion symmetry \cite{Liu2019Topologically-PCCP}. The corresponding LG 69 ($p3m1$) allows the existence of CNPs according to our analysis. We calculated the electronic band structure for the Bi$_2$O monolayer under SOC as plotted in Fig.~\ref{sfig:Bi2O}. One finds that there is a CNP with cubic energy dispersion near the Fermi energy at $\Gamma$ point, more clearly observed in Fig.~\ref{sfig:Bi2O}(e). The CNP is guaranteed by the IRREP $\{^{1}\bar{E}, ^{2}\bar{E}\}$ of the $C_{3v}$ double point group, consistent with our symmetry analysis. The bending-upward energy dispersion is due to the dominant effect of $\omega_1 k^2$ term.

\section{Computational methods}
The first-principles calculations for material candidates were performed based on the density functional theory (DFT), as implemented in the Vienna \emph{ab initio} simulation package (VASP) \cite{Kresse1996Efficiency-CMS,Kresse1996Efficiency-PRB} with the projector augmented wave (PAW) method \cite{Blochl1994Projector-PRB}. The generalized gradient approximation (GGA) with the Perdew-Burke-Ernzerhof (PBE) \cite{Perdew1996Generalized-PRL} realization was adopted for the exchange-correlation potential. The BZ was sampled by the $k$ grids with a spacing of $2\pi \times 0.02$ \AA$^{-1}$ within a $\Gamma$-centered sampling scheme. The crystal structure for all candidate materials were optimized until the forces on the ions were less than $10^{-3}$~eV/\text{\AA}, and the energy convergence criteria were set to be $10^{-8}$~eV. For Ti$_3$C$_2$F$_2$, the optimized lattice constant is $a = 3.078$ \AA, which are close to the experimental values \cite{Naguib2011Two-AM}. The optimized lattice constant of V$_2$TiC$_2$F$_2$ is $a = 2.893$ \AA. For the Bi$_2$O monolayer, the optimized lattice constant is obtained as $a = 4.673$ \AA. To avoid artificial interactions between periodic images, a vacuum layer with the thickness of 15 \AA ~was taken in the calculations.

The numerical results of the lattice models were performed using the open source software Pybinding \cite{Moldovan2020pybinding-Z}, including the band structures and the spectrum of ribbon geometry.

%